\PassOptionsToPackage{table,dvipsnames}{xcolor}
\documentclass[conference,10pt]{IEEEtran}

\usepackage{amsmath,amssymb}
\usepackage{xifthen}
\usepackage{mathtools}
\usepackage{enumerate}
\usepackage{microtype}
\usepackage{xspace}
\usepackage{bm}
\usepackage[T1]{fontenc}
\usepackage{fancyhdr}
\usepackage{lastpage}
\usepackage{bbm}

\newcommand{\mbs}[1]{\pmb{#1}}
\newcommand{\vect}[1]{{\lowercase{\mbs{#1}}}}
\newcommand{\mat}[1]{{\uppercase{\mbs{#1}}}}

\newcommand{\T}{{\scriptscriptstyle\mathsf{T}}}

\renewcommand{\Re}[1][]{\ifthenelse{\isempty{#1}}{\operatorname{Re}}{\operatorname{Re}\left(#1\right)}}
\renewcommand{\Im}[1][]{\ifthenelse{\isempty{#1}}{\operatorname{Im}}{\operatorname{Im}\left(#1\right)}}

\newcommand{\av}{\vect{a}}

\newcommand{\ev}{\vect{e}}

\newcommand{\ellv}{\vect{\ell}}

\newcommand{\Ar}{\mathrm{A}}
\newcommand{\Br}{\mathrm{B}}

\newcommand{\Er}{\mathrm{E}}

\newcommand{\Nr}{\mathrm{N}}

\newcommand{\Rr}{\mathrm{R}}

\newcommand{\Ur}{\mathrm{U}}

\newcommand{\kappav}{\vect{\kappa}}

\newcommand{\piv}{\vect{\pi}}
\newcommand{\tauv}{\vect{\tau}}

\newcommand{\rhov}{\vect{\rho}}

\newcommand{\Pim}{\pmb{\Pi}}

\newcommand{\Lm}{\mat{l}}

\newcommand{\Xm}{\mat{x}}
\newcommand{\Ym}{\mat{y}}
\newcommand{\Zm}{\mat{z}}

\newcommand{\Lc}{{\mathcal L}}

\newcommand{\Nc}{{\mathcal N}}

\newcommand{\Uc}{{\mathcal U}}

\newcommand{\EE}{\mathbb{E}}

\newcommand{\RR}{\mathbb{R}}

\newcommand{\Is}{\mat{\mathsf{I}}}
\newcommand{\Ts}{\mat{\mathsf{T}}}

\newcommand{\Ms}{\mat{\mathsf{M}}}

\newcommand{\CN}[1][]{\ifthenelse{\isempty{#1}}{\mathcal{N}_{\mathbb{C}}}{\mathcal{N}_{\mathbb{C}}\left(#1\right)}}

\renewcommand{\P}[1][]{\ifthenelse{\isempty{#1}}{\mathbb{P}}{\mathbb{P}\left(#1\right)}}
\newcommand{\E}[1][]{\ifthenelse{\isempty{#1}}{\mathbb{E}}{\mathbb{E}\left[#1\right]}}
\newcommand{\I}[1][]{\ifthenelse{\isempty{#1}}{\mathbb{I}}{\mathbb{I}\left\{#1\right\}}}
\renewcommand{\det}[1][]{\ifthenelse{\isempty{#1}}{\mathrm{det}}{\mathrm{det}\left(#1\right)}}
\newcommand{\trace}[1][]{\ifthenelse{\isempty{#1}}{{\rm tr}}{\mathrm{tr}\left(#1\right)}}
\newcommand{\rank}[1][]{\ifthenelse{\isempty{#1}}{\mathrm{rank}}{\mathrm{rank}\left(#1\right)}}
\newcommand{\diag}[1][]{\ifthenelse{\isempty{#1}}{\mathrm{diag}}{\mathrm{diag}\left(#1\right)}}
\newcommand{\Cov}[1][]{\ifthenelse{\isempty{#1}}{\mathsf{Cov}}{\mathsf{Cov}\left(#1\right)}}

\newcommand{\defeq}{\triangleq}

\newtheorem{theorem}{Theorem}%[section]
\newtheorem{simplification}{Simplification}%[section]

\newcounter{enumi_saved}
\usepackage{answers}
\Newassociation{solution}{Solution}{solutionfile}

\AtBeginDocument{\Opensolutionfile{solutionfile}[\jobname]}
\AtEndDocument{\Closesolutionfile{solutionfile}\clearpage
}

\IfFileExists{MinionPro.sty}{
}{
}

\usepackage{subcaption}

\usepackage{graphicx,psfrag}
\usepackage{multirow}
\usepackage{array}
\usepackage{setspace}
\usepackage{bbm,mathtools}
\usepackage{amssymb}
\usepackage{color}
\usepackage{enumitem}
\usepackage{stfloats}
\usepackage[flushleft]{threeparttable}
\usepackage{pgfplots}
\usepackage{booktabs}
\usepackage{makecell}
\usepackage{cite}
\usepackage[table,dvipsnames]{xcolor}
\usepackage{xcolor}
\pgfplotsset{minor grid style={dotted}}
\pgfplotsset{major grid style={dashed}}
\usepackage{tikz}
\usetikzlibrary{fit,positioning,arrows,shapes,shapes.multipart,calc,arrows.meta,shapes.geometric,shapes.misc}
\usetikzlibrary {decorations.fractals,spy} 
\usetikzlibrary{decorations.pathreplacing,angles,quotes,calligraphy}
\usetikzlibrary{automata,fit,positioning,arrows,shapes,shapes.multipart,calc,arrows.meta}
\usepackage[ruled,vlined,linesnumbered]{algorithm2e}

\usepackage[toc,acronym]{glossaries}

\newacronym{AWGN}{AWGN}{additive white Gaussian noise}
\newacronym{MAC}{MAC}{multiple access channel}
\newacronym{UMRA}{UMRA}{unsourced massive random access}
\newacronym{SIMO}{SIMO}{single-input multiple-output}
\newacronym{SISO}{SIMO}{single-input single-output}
\newacronym{iid}{i.i.d.}{independent and identically distributed}
\newacronym{ML}{ML}{maximum likelihood}
\newacronym{PEP}{PEP}{pair-wise error probability}
\newacronym{LLR}{LLR}{log-likelihood ratio}
\newacronym{SNR}{SNR}{signal-to-noise ratio}
\newacronym{SINR}{SINR}{signal-to-interference-plus-noise ratio}

\newacronym{AoI}{AoI}{age of information}
\newacronym{AoII}{AoII}{age of incorrect information}
\newacronym{AVP}{AVP}{age-violation probability}
\newacronym{PMF}{PMF}{probability mass function}
\newacronym{CDF}{CDF}{cummulative distribution function}
\newacronym{CCDF}{CCDF}{complementary cummulative distribution function}
\newacronym{SA}{SA}{slotted ALOHA}
\newacronym{IRSA}{IRSA}{irregular repetition slotted ALOHA}
\newacronym{SIC}{SIC}{successive interference cancellation}
\newacronym{PLR}{PLR}{packet loss rate}
\newacronym{DE}{DE}{density evolution}
\newacronym{IoT}{IoT}{Internet of Things}
\newacronym{EH}{EH}{energy harvesting}
\newacronym{CP}{CP}{contention period}

\newacronym{wp}{w.p.}{with probability}
\newacronym{wrt}{w.r.t.}{with respect to}

\newacronym{BEU}{BEU}{best-effort uniform}
\newacronym{TFB}{TFB}{transmitting only with full battery}

\newacronym{WED}{WED}{wrong-estimate duration}
\newacronym{CED}{CED}{correct-estimate duration}
\newacronym{MEP}{MEP}{missed-event probability}

\newacronym{MDP}{MDP}{Markov decision process}
\newacronym{RL}{RL}{reinforcement learning}
\makeindex
\allowdisplaybreaks

\usepackage{grffile}
\pgfplotsset{compat=newest}
\usetikzlibrary{plotmarks}
\usetikzlibrary{arrows.meta}
\usepgfplotslibrary{patchplots}

\makeatletter
\DeclareRobustCommand{\cev}[1]{%
	{\mathpalette\do@cev{#1}}%
}
\newcommand{\do@cev}[2]{%
	\vbox{\offinterlineskip
		\sbox\z@{$\m@th#1 x$}%
		\ialign{##\cr
			\hidewidth\reflectbox{$\m@th#1\vec{}\mkern4mu$}\hidewidth\cr
			\noalign{\kern-\ht\z@}
			$\m@th#1#2$\cr
		}%
	}%
}
\makeatother

\newcommand{\Punnoticed}[1]{\rho_{#1}}
\newcommand{\Poblivious}[1]{\kappa_{#1}}

\newcommand{\of}[1]{^{(#1)}}

\newcommand{\ind}[1]{{\mathbbm{1}{\{#1\}}}}

\renewcommand{\E}[2][]{\mathbb{E}_{#1}\!\left[#2\right]}
\renewcommand{\P}[2][]{\mathbb{P}_{#1}\!\left[#2\right]}

\newcommand{\sub}[1]{_{\mathrm{#1}}}
\newcommand{\sups}[1]{^{\mathrm{#1}}}

\renewcommand{\defeq}{=}

\newcommand{\Pmiss}{P_{\rm ME}}

\newcommand{\revise}[1]{{#1}}

\makeatletter
\newcommand{\removelatexerror}{\let\@latex@error\@gobble}
\makeatother

\title{Information Age and Correctness for Energy Harvesting Devices with Random Access}

\author{\IEEEauthorblockN{Khac-Hoang Ngo\IEEEauthorrefmark{1}, Giuseppe Durisi\IEEEauthorrefmark{2}, Petar Popovski\IEEEauthorrefmark{3}}
\IEEEauthorblockA{\IEEEauthorrefmark{1}Department of Electrical Engineering, Linköping University, 58183 Linköping, Sweden}
\IEEEauthorblockA{\IEEEauthorrefmark{2}Department of Electrical Engineering, Chalmers University of Technology, 41296 Gothenburg, Sweden}
\IEEEauthorblockA{\IEEEauthorrefmark{3}
Department of Electronic Systems, Aalborg University, 9220 Aalborg Øst, Denmark}
\thanks{This work was supported in part by ELLIIT, by the Swedish research council under grant 2021-04970, and by the Velux Foundation, Denmark, through the Villum Investigator Grant WATER, nr. 37793.}
\vspace{-.8cm}
}

\IEEEoverridecommandlockouts
\begin{document}
	
	\maketitle
	\begin{abstract} 
        \revise{We investigate accuracy and freshness of status updates from a large number of energy-harvesting devices that monitor two-state Markov processes %and send status updates to a gateway using 
        and access the medium using the slotted ALOHA protocol without feedback.} %The process state evolves according to a two-state Markov chain. 
        %We let the devices adjust their transmission probabilities according to their process state transitions and current battery levels. 
        Using a Markovian framework, we analyze the average value of a {generic} state-dependent penalty function that grows whenever there is a state estimation error. {The
        age of incorrect information (AoII) is an example of such penalty function.} %, as well as the average value of a state-dependent penalty function that grows with the AoII. 
        We propose an accurate and easy-to-compute approximation for the average penalty. Numerical results demonstrate the benefits of optimizing the transmission probabilities \revise{according to the process state transitions and current battery levels} to minimize the average penalty. {Minimizing a state-independent penalty function} can be highly suboptimal in terms of average penalty when one of the process states is critical, i.e., entails a high penalty if wrongly estimated. %{highlight that the AoII fails to capture the varying significance of different state estimation errors, which can be achieved with a suitable penalty function.} 
        Furthermore, minimizing the average penalty does not guarantee a low probability of misdetecting a critical state period.
        
	\end{abstract}
	% \begin{IEEEkeywords}
	% 	Internet of Things, random access, slotted ALOHA, age of information, energy harvesting
	% \end{IEEEkeywords}

    % \vspace{-.3cm}
	%-----------------------------------------------
	\section{Introduction} \label{sec:intro}

    \revise{One of the next challenges in the development of \gls{IoT} systems is to support ultra-low-complexity and ultra-low-power devices with extended lifespans and that do not require manual battery replacement or recharging~\cite{Butt24_ambientIoT}. These devices harvest energy from ambient sources, such as radio waves, light, motion, and heat~\cite{Kamalinejad2015wireless}. They perform remote monitoring of physical processes, and report observations %of these processes 
    to a central gateway for analysis and decision-making. This paper addresses methods to ensure both \textit{accurate} and \textit{fresh} status updates from a large number of energy-harvesting devices. 
    
    We assume that the devices access the medium following the slotted ALOHA protocol, which underpins many modern uncoordinated access protocols. %Many existing commercial solutions~\cite{ETSI2020DVB,LoRa,SigFox} employ simple uncoordinated medium access protocols based on variations of the classic ALOHA scheme. %~\cite{Abramson1970ALOHA}. 
    To capture both the accuracy and freshness of the status updates from the devices, 
    we consider the \gls{AoII} metric~\cite{Maatouk20_AoII}, which generalizes the \gls{AoI} metric~\cite{Yates2021AoI} by capturing the informativeness of the successfully received updates. %, the \gls{AoII} measures the time elapsed since the last moment when the representation of the process state at the gateway becomes inaccurate. %, thereby capturing both the freshness and accuracy of the updates. 
    %This dual focus on timeliness and detection error has made \gls{AoII} a compelling metric. %, inspiring a range of studies. 
    Existing \gls{AoII} studies have primarily explored scenarios involving a single device (e.g.,~\cite{%Joshi21,Bountrogiannis24,
    Chen24}) or multiple coordinated devices (e.g.,~\cite{%Kriouile21,
    Ayik23}) with a stable power supply. These studies leverage slot-wise feedback from the gateway to enable the device(s) or a central scheduler to track the current \gls{AoII} and plan update transmissions accordingly. 
    
    %Traditional performance metrics for these protocols, such as packet loss rate and throughput, capture only the accuracy/reliability of the transmitted updates. {This has motivated the introduction of new metrics such as the} \gls{AoI}~\cite{Yates2021AoI}, which focuses on freshness by measuring the time elapsed since the generation of the last update available at the gateway. While \gls{AoI} has gained popularity, it falls short in scenarios where the accuracy of the monitored data is critical. 

    The behavior of uncoordinated medium access protocols through the lens of \gls{AoII} remains, however, {less} explored.  
    An early effort is~\cite{Nayak23}, which studied \gls{AoII} in slotted ALOHA systems where the devices exploit feedback to adjust access probabilities and prioritize processes with higher penalties. {In~\cite{Chiariotti2024distributed}, each device listens to feedback intended for all devices and applies dynamic epistemic logic to avoid collisions and reduce \gls{AoII}.} For the case of slotted ALOHA without feedback, the author of~\cite{Munari24_AoII} derives a closed-form expression of the average \gls{AoII}. These studies, however, do not address energy-harvesting devices. While the influence of energy harvesting on the \gls{AoI} in random access networks has been explored in~\cite{Sleem2020,Ngo24statusupdate}, its impact on the \gls{AoII} has not been determined.
    }

\begin{figure}
        \centering
        \scalebox{.85}{
        \begin{tikzpicture}
            \tikzset{node distance=3cm, % Minimum distance between two nodes. Change if necessary.
                every state/.style={ % Sets the properties for each state
                    semithick,
                    fill=gray!10},
                initial text={},     % No label on start arrow
                double distance=4pt, % Adjust appearance of accept states
                every edge/.style={  % Sets the properties for each transition
                    draw,
                    ->,>=stealth',     % Makes edges directed with bold arrowheads
                    auto,
                    semithick}}
                \node[state] (S0) {$0$};
                \node[state, right of= S0] (S1) {$1$};
                
                \draw (S0) edge[loop left] node[left,midway] {$q_{00}$} (S0);
                \draw (S1) edge[bend left = 20] node[below,midway] {$q_{10}$} (S0);
                \draw (S1) edge[loop right] node[right,midway,align=center] {$q_{11}$} (S1);
                
                \draw (S0) edge[bend left = 20] node[above,midway] {$q_{01}$} (S1);
        \end{tikzpicture}
        }
        \vspace{-.2cm}
        \caption{The Markov process tracked by a device.}
        \label{fig:source}
        \vspace{-.6cm}
    \end{figure}
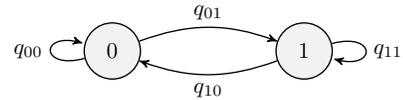

    In this paper, we analyze the \gls{AoII} in a slotted ALOHA network where each energy-harvesting device monitors the two-state Markov process {depicted in Fig.~\ref{fig:source}. This process is relevant in systems with binary states, such as the active/idle status of a machine, and occupied/unoccupied status of a resource unit.
    In a slot, the process moves from state $i$ to state $j$ with probability~$q_{ij}$ for $i,j \in \{0,1\}$.} % and relies on harvested energy for update transmission.
    We assume no feedback from the gateway. %To adapt the protocol to this scenario, 
    We let the devices adjust their transmission probabilities based on the process state transitions and their current battery levels. %Using a Markovian approach, we analyze the average \gls{AoII}. 
    We also go beyond the AoII and consider a {state-dependent} penalty function that increases with the \gls{AoII}. {We assume that {state $1$} is critical, i.e., a wrong estimation of {state $1$} entails a high loss to the system. We therefore associate a larger penalty with this state.} % to capture the significance of different state estimation errors.} 
    We derive the average penalty for penalty functions that grow linearly or as a power of the \gls{AoII}. Furthermore, leveraging an absorbing Markov chain analysis {similar to~\cite{Ngo24statusupdate}}, we propose an efficient and accurate approximation for the average penalty and for the probability of misdetecting a period with {the} critical state, called the \gls{MEP}.

    In numerical experiments, we assume that each slot comprises multiple uses of an \gls{AWGN} channel, {which is relevant in systems where the devices estimate their channel based on downlink pilot, broadcast from the gateway, and pre-equalize their uplink signal}. %\pp{PP: We need to justify AWGN channel in a wireless setup and the usual way is with channel inversion based on a pilot sent from the BS. Note that we need to define this consistently, such that if the channel is too weak, then the device needs to use excessive energy and therefore it stays silent; so, the activity of the device is modulated by its content (data), battery level AND channel quality.}  
    We evaluate three update strategies: a \textit{reactive} strategy, where the devices transmit only upon detecting a process state change {(i.e., $0 \to 1$ or $1\to 0$)}; a \textit{random} strategy, where the devices transmit regardless of the process state; and a \textit{hybrid} strategy that adapts the transmission probability to all four possible state transitions. %\pp{PP: How do we know at this point that there are four? Needs to be related to the device model. I suggest to have a better description of the process/state transitions and explain the problem+contribution on those. Even a small figure would help possible process state transitions and also clarify the three strategies studied}}. %Our results reveal a tradeoff among these strategies. 
    Overall, the hybrid strategy achieves the lowest penalty, and its advantage is more pronounced when the transition probabilities are asymmetric, {i.e., when $q_{10} \ne q_{01}$}. We shall refer to the corresponding process as asymmetric process. The reactive strategy incurs a high penalty for infrequent state changes, but can match the hybrid strategy's performance when the processes change state rapidly. This strategy also achieves a low \gls{MEP} by conserving energy during periods of no state change. The random strategy fails to prioritize certain state changes, leading to poor performance for asymmetric processes with high state transition rates. For asymmetric processes, we also demonstrate that optimizing for a state-independent penalty function can be highly suboptimal {with respect to a penalty function that penalizes more significantly a wrong estimation of the critical state. %Indeed, the \gls{AoII} overlooks the importance of critical state estimation errors}.
    This underscores the necessity of tailoring the penalty function to the significance of each state.

	% \subsection{Notation} 
	%The remainder of the paper is organized as follows. In Section~\ref{sec:model}, we present the system model and the \gls{AoI} metrics. In Section~\ref{sec:preliminaries}, we present some preliminary results that will be used in our \gls{AoI} analysis. We then provide an exact and approximate \gls{AoI} analysis in Sections~\ref{sec:AoII} and~\ref{sec:AoI_approx}, respectively. In Section~\ref{sec:results}, we present numerical results and discussions. We conclude the paper in Section~\ref{sec:conclusions}. The proofs are deferred to the appendix.
    
	\subsubsection*{Notation}
	We denote system parameters and constants by uppercase non-italic letters, e.g.,~$\Ur$, or Greek letters. We denote scalar random variables by uppercase italic letters, e.g.,~$X$, and their realizations by lowercase italic letters, e.g.,~$x$. Column vectors are denoted likewise with boldface letters. %, e.g., a random vector $\Xm$ and its realization~$\xv$. %All vectors are column vectors. 
    %The superscript~$^\T$ denotes the transpose. 
    We use sans-serif, uppercase, and boldface letters, e.g.,~$\Ms$, to denote deterministic matrices. 
	By $\Is$ %, $\mathbf{0}_{m}$, 
    and $\mathbf{1}$, we denote the %$m\times m$ 
    identity matrix and %$m\times 1$ all-zero matrix, and $m\times 1$ 
    the all-one vector, respectively. %, whose dimensions are clear from the context. %We also write $\mathbf{0}_{m \times 1}$ as $\mathbf{0}_{m}$. 
	%The first entry of vector $\xv$ is denoted by $[\xv]_1$; 
	%The diagonal matrix with diagonal elements $(d_1,d_2,\dots,d_m)$ is denoted by $\diag(d_1,d_2,\dots,d_m)$. 
    We denote by $\ind{\cdot}$ the indicator function; $[m:n] = \{m,m+1,\dots,n\}$; and %to denote the set of integers from $m$ to $\Nr$, and 
	$[n] = [1:n]$. %, and $x^+ = \max\{0,x\}$.  We denote %the Binomial distribution with parameters $(n,p)$ by $\mathrm{Bino}(n,p)$, %, and its probability density function evaluated at~$k$ as $\mathrm{Bino}(k;n,p) \triangleq \binom{\Nr}{k}p^k(1-p)^{n-k}$. 
	%We denote by ${\rm Mult}(n,k,\{p_i\}_{i=1}^k)$ 
	%and 
	We denote the multinomial distribution with $n$ trials, $k$ events, and event probabilities $\{p_i\}_{i = 1}^k$ by ${\rm Mul}(n,k,\{p_i\}_{i=1}^k)$. %, and the geometric distribution with success probability $p$ by ${\rm Geo}(p)$. %For a Markov chain $M(t)$, we denote the transition probability by $\P{m \to m'} = \P{M(t+1) = m' \vert M(t) = m}$ and the time-reversed transition probability by $\P{m \leftarrow m'} = \P{M(t) = m \vert M(t+1) = m'}$. 
	%The transpose and Moose-Penrose inverse of a real matrix $\Am$ are denoted by $\Am^\T$ and $\Am^\dagger = \Am^\T(\Am\Am^\T)^{-1}$, respectively. 
	%The logarithm of an $m\times m$ matrix $\Am$ is defined as $\ln \Am = \sum_{i=1}^\infty (-1)^{i+1} \frac{(\Am - \Is_m)^i}{i}$ whenever the series converges~\cite[Def.~2.6]{Hall2003_Lie}. 
	
	% \subsection{Reproducible Research} 
	% The Matlab code used for our numerical results is available at: \hoang{I will add the link.}

        % \clearpage
	%-----------------------------------------------
	\section{System Model} \label{sec:model}
	We consider a system with $\Ur$ devices %attempting to 
    delivering time-stamped status updates (also called packets) about distributed processes to an \gls{IoT} gateway through a shared wireless channel. %We assume that each device receives readings from a sensor, and thus cannot generate fresh updates at will. Updates are generated independently across sensors. 
	Time is divided into slots and the devices are assumed slot synchronous. Without loss of generality, we let the slot length be $1$ and assume that each update transmission spans a slot. 
	
	\subsubsection{Process Evolution}
	Each device tracks a process described by a %\gls{iid} 
    two-state discrete-time Markov chain {illustrated in Fig.~\ref{fig:source}.} 
    Without loss of generality, we focus on a specific device and denote the state of the corresponding Markov chain at slot $n\in \mathbb{N}$ as $X\of{n} \in \{0,1\}$. We assume that the state $X\of{n}$ can change value at the beginning of each slot. We denote by $q_{ij}$ the probability of transition from state $i$ to state $j$ for $i,j \in \{0,1\}$. %The average total number of transitions per slot is $\bar{q} = 2 q_{01} q_{10}/(q_{01} + q_{10})$.

	\subsubsection{Energy Harvesting}
	Each device has a rechargeable battery with a capacity of $\Er$ energy units, and harvests ambient energy to recharge it. %
	As in~\cite{Ibrahim2016,Bae2017,Demirhan2019,Chen2022AoI,Ngo24statusupdate}, we model energy harvesting as an independent Bernoulli process. Specifically, a device obtains a new energy unit in each slot with a given probability called the energy harvesting rate. The energy harvesting process is independent across slots and across devices. If the battery is full, the device pauses harvesting. 
    
    In certain scenarios, the monitored processes can also be the processes that deliver energy for harvesting. For example, vibration sensors on a bridge can report the traffic flow while harvesting energy from the same vibration. A heavy traffic flow is critical to report and also increases the energy harvesting rate. To accommodate these scenarios, we let the energy harvesting rate vary with the process state.
	In a slot, a device with process state $x \in \{0,1\}$ has energy harvesting rate~$\gamma_x > 0$. %We refer to $\gamma$ as the energy harvesting rate.
	 %A device can transmit a packet if its battery is not depleted. 
	%We denote by $\nu_b$ (calculated in Section~\ref{sec:bat_generic_device}) the steady-state probability that the battery level of an arbitrary device is $b \in [0:\Er]$. %We shall compute $\{\nu_b\}$ in Section~\ref{sec:steady_state}. %In steady-state, the numbers of devices having battery level $e \in [0:\Er]$, $\{U_e\}_{e=0}^\Er$, follow the ${\rm Mult}(U-1,B+1,\{\nu_e\}_{e = 0}^\Er)$ distribution. 

    We refer to the pair $(x,b)$ of process state $x$ and battery level $b$ of a device in a slot as the \emph{{process-battery}} state. We characterize the process-battery profile of the other $\Ur - 1$ devices by the vector $\Lm = (L_{0,0}, \dots, L_{(0,\Er)}, L_{(1,\Er)},\dots, L_{(1,\Er)})$ containing the number $L_{(x,b)}$ of devices having process-battery state $(x,b)$ for $x \in \{0,1\}$ and $b\in [0:\Er]$ among these devices.

	\subsubsection{Medium Access Protocol}  \label{sec:protocol}
	The devices access the medium following slotted ALOHA. %, adapted to our scenario. %~\cite{Roberts1972}. 
    %To adapt %the protocol 
    %to our scenario, 
    We let each device choose its transmission probability according to its battery level and process state transition. 
	Specifically, consider a device with battery level~$b$ whose state changes from~$i$ in the previous slot to~$j$ in the current slot. We let this device transmit an update of the current %process 
    state with probability $\pi_{b}\of{ij}$, using all available energy. % within a single slot. % If its state has changed in a slot, it transmits a status update with probability $\pi_{b}$. Otherwise, it transmits a status update with probability $\chi_b$. 
    %Whenever the device transmits, it uses all available energy within a single slot.  %\footnote{Our analysis can be extended to the case where the devices retransmit the latest update whenever they do not have a new update. We shall discuss this case in details in Section~\ref{sec:retx}.} %Otherwise, it retransmits the latest update using  $b_{\rm t}$ energy units with probability $\hat x_{b,b_{\rm t}}$. 
	Obviously, $\pi_{0}\of{ij} = 0$ for all $i$ and $j$. The matrix $\Pim = [\piv\of{00} \ \piv\of{01} \ \piv\of{10} \ \piv\of{11}]$ with $\piv\of{ij} = [\pi_1\of{ij}\ \pi_2\of{ij} \dots \pi_\Er\of{ij}]^{\T}$ contains the design parameters of the protocol. %For convenience, we denote the probability that a device with battery level $b$ accesses the channel by
	%\begin{align}   
	%$\rho_{b} = \sum_{i,j \in \{0,1\}} \alpha_i q_{ij} \pi_{b}\of{ij}$.
	We consider the case without feedback from the gateway. %, so the device does not know which of its updates was received successfully. %\pp{PP: So, the device has no way to know which of its updates was received successfully in the past?}

    {
    The strategy $\Pim$ is  optimized offline and then fixed during device operation.\footnote{If there is feedback, $\Pim$ should be dynamically adjusted based on, e.g., an estimate of the current process-battery profile of the other devices.}
    \!} We shall examine three strategies: i) a \emph{reactive} strategy, where the device only transmits when there is a process state change, i.e., $\pi_b\of{ij} = 0$ if $i = j$; ii) a \emph{random} strategy, where the device uses the same transmission probability regardless of its process state, i.e., $\pi_b\of{ij} = \pi_b$ for every $(i,j)$; iii) a \emph{hybrid} strategy, where the transmission probability $\pi_b\of{ij}$ can be chosen between $0$ and $1$ for every $(i,j,b)$. These strategies were studied in~\cite{Munari24_AoII} for a setting with unlimited energy, unlike our energy-constrained framework.
    
    We denote by $\omega_{b,\Lm}$ the probability that an update transmitted with~$b$ energy units is correctly decoded given that the remaining $\Ur-1$ devices have process-battery profile $\Lm$. 
	%The functional dependency of $\omega_{b,\Lm}$ on $(b,\Lm)$ captures the impact of the transmit power and the interference from the other devices on the decoding of the current device. All analytical results in the paper hold for general $\omega_{b,\Lm}$, while in numerical experiments in Section~\ref{sec:results}, we shall instantiate $\omega_{b,\Lm}$ by considering a real-valued \gls{AWGN} channel.
 %    At steady state, the average successful decoding probability of an update transmitted with $b$ energy units is
	% \begin{equation} \label{eq:avg_w}
	% 	\bar{\omega}_b = \E{\omega_{b,\Lm}},
	% \end{equation}
	% where the expectation is over the steady-state distribution %$ follows the multinomial distribution with number of trials $\Ur-1$, number of events $B+1$, and event probabilities $\{\nu_i\}_{i = 0}^\Er$, denoted by $
	% ${\rm Mult}(\Ur-1,2\Er+2,\{\nu_{(x,b)}\}_{x \in\{0,1\}, b \in [0:\Er]})$ of $\Lm$. Here, $\{\nu_{(x,b)}\}_{x\in\{0,1\}, b\in[0:\Er]}$, computed in Section~\ref{sec:source_battery_evolution}, is the steady-state distribution of the process-battery state. 
    %The average throughput, i.e., the average number of packets decoded per slot, is given by
	%\begin{align}
	%\Tr &= \Ur \sum_{x' \in\{0,1\}, b' \in [0:\Er]} \nu_{(x',b')} \sum_{x \in\{0,1\}}  q_{x'x} \pi\of{x'x}_{b'} \bar{\omega}_{b'}.%     \E[\{U_e\}_{e=0}^\Er]{\sum_{e = 0}^\Er \alpha U_e \pi_e \omega_e}. 
	%    \label{eq:throughput}
	%\end{align}
	%where the numbers of devices having battery level $e \in [0:\Er]$, $\{U_e\}_{e=0}^\Er$, follow the ${\rm Mult}(U-1,B+1,\{\nu_e\}_{e = 0}^\Er)$ distribution. 
	
	\subsubsection{Performance Metrics} \label{sec:performance_metrics}
    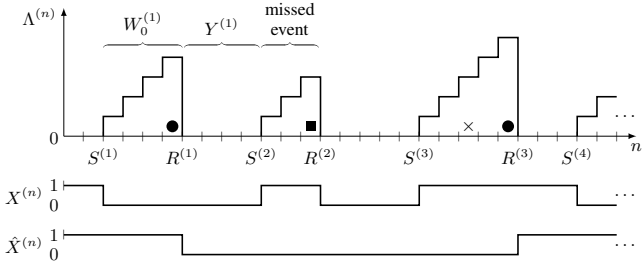
\begin{figure}[t!]
		\centering
		\scalebox{.75}{\begin{tikzpicture}[xscale=0.35,yscale=0.35,domain=0:25,samples=400]
				\tikzstyle{every node}=[font=\small]
				
				\draw[-latex] (1,0) -- (30,0) node[below] {$n$};
				\draw[-latex] (1,0) -- (1,6.5) node[left,yshift=-.15cm] {$\Lambda\of{n}$};
				
				\foreach \p in {2,3,4,5,6,7,8,9,10,11,12,13,14,15,16,17,18,19,20,21,22,23,24,25,26,27,28,29}{\node at (\p,0) () {\tiny $|$};}
				
				\draw[thick] (3,0) -- (3,1) -- (4,1) -- (4,2) -- (5,2) -- (5,3) -- (6,3) -- (6,4) -- (7,4) -- (7,0);

				\draw[thick] (11,0) -- (11,1) -- (12,1) -- (12,2) -- (13,2) -- (13,3) -- (14,3) -- (14,0);
								
				\draw[thick] (19,0) -- (19,1) -- (20,1) -- (20,2) -- (21,2) -- (21,3) -- (22,3) -- (22,4) -- (23,4) -- (23,5) -- (24,5) -- (24,0);
				
				\draw[thick] (27,0) -- (27,1) -- (28,1) -- (28,2) -- (29,2);
				
				\node at (29.5,1) () {$\dots$};
%				\node at (-.5,1.1) () {\small $1$};
				\node at (.5,-0.1) () {$0$};
%				\node at (-.5,7) () {$\theta$};
%				\draw[dashed] (0,7) -- (21,7);
				
				\node at (3,-1) () {$S\of{1}$};
				\node at (7,-1) () {$R\of{1}$};
				\node at (11,-1) () {$S\of{2}$};
				\node at (14,-1) () {$R\of{2}$};
				\node at (19,-1) () {$S\of{3}$};
				\node at (24,-1) () {$R\of{3}$};
				\node at (27,-1) () {$S\of{4}$};

                \draw[thick] (1,-2.5) -- (3,-2.5) -- (3,-3.5) -- (11,-3.5) -- (11,-2.5) -- (14,-2.5) -- (14,-3.5) -- (19,-3.5) -- (19,-2.5) -- (27,-2.5) -- (27,-3.5) -- (29,-3.5);
                \node at (29.5,-3) () {$\dots$};
                \node at (1,-2.5) () {\tiny $|$};
                \node at (.5,-2.5) () {$1$};
                \node at (.5,-3.5) () {$0$};
                \node at (-1,-3) () {$X\of{n}$};
                
                \draw[thick] (1,-5) -- (7,-5) -- (7,-6) --  (24,-6) -- (24,-5) -- (29,-5);
                \node at (29.5,-5.5) () {$\dots$};
                \node at (1,-5) () {\tiny $|$};
                \node at (.5,-5) () {$1$};
                \node at (.5,-6) () {$0$};
                \node at (-1,-5.5) () {$\hat X\of{n}$};
				
				\node[circle,fill=black,inner sep=0pt,minimum size=6pt] at (6.5,.5) () {};
				
				\node[circle,fill=black,inner sep=0pt,minimum size=6pt] at (23.5,.5) () {};
				
				\node[rectangle,fill=black,inner sep=0pt,minimum width=5pt,minimum height=5pt] at (13.5,.5) () {};
				
				\node at (21.5,.5) () {$\bf \times$};
				
				\draw [decorate, decoration = {calligraphic brace}] (3,4.5) -- node[above=.1cm,midway] {$W_0\of{1}$} (7,4.5);
				
				\draw [decorate, decoration = {calligraphic brace}] (7.1,4.5) -- node[above=.1cm,midway] {$Y\of{1}$} (10.9,4.5);
				
				\draw [decorate, decoration = {calligraphic brace}] (11,4.5) -- node[above=.1cm,midway,align=center] {missed \\ event} (13.9,4.5);
				
%				\draw [decorate, decoration = {calligraphic brace}] (17,6.7) -- node[below=.2cm,pos=.8] {\footnotesize $Y\!\!-\!\theta \!+\! 1$} (15,6.7);
			\end{tikzpicture}
		}
        
        \vspace{-.1cm}
		\caption{Example evolution of the \gls{AoII} over time. A circle denotes a slot in which an update is successfully delivered, resetting the \gls{AoII} to $0$. A cross denotes a failed update delivery. A square denotes a slot in which the metric is reset after an unnotified state change. {The quantities $W_x\of{i}$, $Y\of{i}$, $R\of{i}$, and $S\of{i}$ are defined in Section~\ref{sec:AoII}.}}
		\label{fig:AoII_process}
		\vspace{-.52cm}
	\end{figure}
    
    The gateway uses the latest update from the device as an estimate $\hat{X}\of{n}$ of the monitored process $X\of{n}$. That is, $\hat{X}\of{n} = X\of{n}$ if a new update is successfully received from the device in slot $n$; otherwise, $\hat{X}\of{n} = \hat{X}\of{n-1}$. {The \gls{AoII} of a generic device is defined as %the time elapsed since the last slot at which the receiver started having a wrong estimate of the source state of that device. It is defined rigorously as %We define the \gls{AoII} of a generic device in slot $s$ as
	% \begin{equation} \label{eq:AoII}
        $\Lambda\of{n} \defeq g\sub{time}(n) g\sub{info}(X\of{n},\hat X\of{n})$ where $g\sub{time}(\cdot)$ and $g\sub{info}(\cdot)$ are time and information penalty functions, respectively~\cite{Maatouk20_AoII}. We consider the case where $g\sub{time}(n) = n - \theta(n)$ with $\theta(n) = \max\{n' \le n \colon \delta\of{n'-1} = 0, \delta\of{n'} = 1\}$ being the latest slot at which the receiver started having a wrong state estimate, and $g\sub{info}(X\of{n},\hat X\of{n})$ is the error indicator 
		$\delta\of{n} = |\hat{X}\of{n} - X\of{n}| \in \{0,1\}$.  
    } 
	The \gls{AoII} process is ergodic and follows an evolution profile exemplified in Fig.~\ref{fig:AoII_process}.
	%This process is ergodic. %We focus on its average value, given by
	% \begin{equation}
	% 	\overline{\Lambda} = \lim\limits_{k \to\infty} \frac{1}{k} \sum_{n=1}^k\Lambda\of{n}. \label{eq:AoII}
	% \end{equation}	
	% and its tail probability, called \gls{AoII} violation probability, i.e., 
	% \begin{equation}
	% 	\zeta(\theta) = \lim\limits_{k \to\infty} \frac{1}{k} \sum_{n=1}^k \ind{\Lambda\of{n} > \theta}.\label{eq:AoII_VP}
	% \end{equation}
	Notice that an \gls{AoII} increase is triggered whenever there is a process state estimation error, %discrepancy between the process state and its estimate, 
    regardless of the current state. In practice, a process state can be critical, e.g., if the loss from taking a wrong action upon missing this state is  high~\cite{Kountouris21}. {To capture this, we need a state-dependent time penalty, for which we use a nondecreasing penalty function of the \gls{AoII} $f_{X\of{n}}(\Lambda\of{n})$ that depends on the current state $X\of{n}$.
    %We remark that in this penalty function, we let the time penalty be state-dependent, as opposed to state-independent time penalty in the \gls{AoII}. For instance,  the general-form \gls{AoII} does not capture 
    We focus on the power penalty function
    %Examples include the linear function
    % \begin{equation}
        %$f_{X\of{n}} = \alpha_{X\of{n}} \Lambda\of{n}$ %, \label{eq:pen_linear} 
    % \end{equation}
    %\revisee{and the power function
    % \begin{equation}
        $f_{X\of{n}} =  (\Lambda\of{n})^{\alpha_{X\of{n}}}$, % \label{eq:pen_power} 
        %\\
        % f_{X\of{n}}(\Lambda\of{n}) &= \exp\big(\alpha_{X\of{n}} \Lambda\of{n}\big) - 1, \label{eq:pen_exp}
    % \end{equation}
    % For example, the penalty function can be defined as 
    % \begin{align} \label{eq:penalty_exp}
    %     f_{X\of{n}}(\Lambda\of{n}) = \exp\big(\alpha_{X\of{n}} \Lambda\of{n}\big) - 1,
    % \end{align}
    where the nonnegative parameters $\alpha_{x}$ represent the significance of an erroneous estimation of the state~$x \in \{0,1\}$.} Here, without loss of generality, we assume that the process state~$1$ is critical, and thus let $\alpha_1 > \alpha_0$. We assume that $f_0(0) = f_1(0) = 0$, i.e., a correct state estimate entails no penalty. As the penalty function captures the relative importance of different error events, it is considered a semantic-aware metric.  We are interested in characterizing the average penalty
    \begin{equation}
		\overline{F} = \lim\limits_{k \to\infty} \frac{1}{k} \sum_{n=1}^k f_{X\of{n}}(\Lambda\of{n}). \label{eq:avg_penalty}
	\end{equation}	
    %{We consider only penalty functions for which $\overline{F}$ is finite.}%Note that $\overline{F}$ can be infinite, and thus nonmeaningful, for some penalty functions.
    %\footnote{{For our setting, $\overline{F}$ is infinite, and thus nonmeaningful, for, e.g., the exponential penalty function $f_{X\of{n}}(\Lambda\of{n}) = \exp\big(\alpha_{X\of{n}} \Lambda\of{n}\big) - 1$.}} 
    The average \gls{AoII}, denoted by $\overline{\Lambda}$, is obtained from~\eqref{eq:avg_penalty} by simply replacing $f_{X\of{n}}(\Lambda\of{n})$ with $\Lambda\of{n}$.
    
    We also consider the \gls{MEP}, which is the probability of misdetecting a period of critical state, i.e., a period for which the process transitions to state~$1$ and leaves it without the receiver noticing. We %call this probability \gls{MEP} and 
    denote this probability by~$\Pmiss$.

	\section{Markov Analysis of the Operation of a Device} \label{sec:preliminaries}
	
	%{We follow a Markov analysis similar to~\cite{Ngo24statusupdate}.} %we analyze the Markov chains that characterize the operation of the devices. 
	
	\subsection{Process-Battery Evolution of a Generic Device} \label{sec:source_battery_evolution}
	Consider a generic device and let $B\of{n}$ be its battery level in slot $n$. Recall that its process state in slot $n$ is denoted by $X\of{n}$. The process-battery state evolution of the device is captured by the Markov chain $(X\of{n},B\of{n})$. % shown in Fig.~\ref{fig:markov_XB}.
    Consider a transition from state $(x',b')$ to state $(x,b)$. This transition requires that the process moves from state $x'$ to state $x$, which occurs with probability $q_{x'x}$. Furthermore, if the device transmits, which occurs with probability $\pi\of{x'x}_{b'}$, the probability that it ends up with battery level $b$ is
    % \begin{equation}
        $\phi\sups{trans.}_{b} = (1-\gamma_x) \ind{b = 0} +  \gamma_x \ind{b=1}$.
    % \end{equation}
    If the device does not transmit, the probability that it moves from battery level $b'$ to $b$ is
    % \begin{equation}
        $\phi_{b' \to b}\sups{no~trans.} = (1-\gamma_x \ind{b' \ne \Er}) \ind{b=b'} + \gamma_x \ind{b=b'+1}$.
    % \end{equation}
    Therefore, using the law of total probability, we obtain the transition probabilities of the chain $(X\of{n}, B\of{n})$ as
    \begin{multline}
        \P{(x',b') \to (x,b)} = \\ q_{x'x}\big[ \pi\of{x'x}_{b'} \phi\sups{trans.}_{b} + (1-\pi\of{x'x}_{b'}) \phi_{b' \to b}\sups{no~trans.}\big]. \label{eq:p_XB}
    \end{multline}
 % Each state represents a battery level. The transition probabilities between the states can be readily computed. Specifically, a device in state $0$ cannot transmit, thus it either remains in this state if it does not harvest energy (with probability $1-\gamma$) or jumps to state $1$ if an energy unit is harvested (with probability $\gamma$). A device in state $i \in [\Er]$ moves to state $0$ if it transmits an update and no energy harvested (with probability $\rho_{i} (1-\gamma)$), and moves to state $1$ if it transmits an update and harvests one energy unit (with probability $\rho_{i} \gamma$). Furthermore, this device remains in state $i$ if it neither transmits nor harvests energy (with probability $(1-\rho_i)(1-\gamma)$), and jumps to state $i+1$ if it does not transmit and an energy unit is harvested  (with probability $(1-\rho_i)\gamma$). Note that if $i = \Er$, the transition $i \to i+1$ is merged into the transition $i \to i$ because of battery overflow. 
	% To summarize, the transition probabilities are given by 
	% \begin{equation}
	% 	\P{i\to j} = \rho_{i}  \phi\sups{trans.}_{j}  +  (1-\rho_{i})\phi_{i \to j}\sups{no~trans.}, \label{eq:p_transition_battery}
	% \end{equation}
	% for $ i,j \in [0:\Er]$, where $$\phi\sups{trans.}_{j} = (1-\gamma) \ind{j = 0} +  \gamma \ind{j=1}$$ is the probability that the device has battery level $j$ after a transmission, and $$\phi_{i \to j}\sups{no~trans.} = (1-\gamma \ind{i \ne \Er}) \ind{j=i} + \gamma \ind{j=i+1}$$ is the probability that the device moves from battery level $i$ to~$j$ while not transmitting. 
	From these transition probabilities, we compute the steady-state distribution $\{\nu_{(x,b)}\}$ by solving the balance equations.

	\subsection{Process-Battery Profile Evolution of $\Ur-1$ Devices} \label{sec:Markov_bat_profile}
    %In a given slot, let $H_{(x,b)}$ be the number of devices, among the $\Ur-1$ devices that are not tracked, that has process-battery state $(x,b)$. 
    %We refer to $\Lm = (H_{(0,0)}, \dots, H_{(0,\Er)}, H_{(1,0)}, \dots, H_{(1,\Er)})$ as the process-battery profile of these $\Ur-1$ devices. It 
    The process-battery profile $\Lm$ of the other devices takes value in the set
	% \begin{align}
	$\Lc = \big\{\{\ell_{(x,b)}\}_{x \in \{0,1\},b\in [0:\Er]} \in [0:\Ur-1]^{2\Er+1}  
    %\notag \\ & \qquad 
    \colon \textstyle \sum_{x = 0}^1 \sum_{b = 0}^\Er \ell_{(x,b)} = \Ur-1\big\}$
	% \end{align}
	with cardinality 
    $|\Lc| = \binom{\Ur+2\Er+1}{2\Er+2}$.
	% We consider a Markov chain where each state represents an element of $\Lc$. 
	%We now describe the evolution of $\Lm$ across slots. 
    Let $\ellv'$ and $\ellv$ be the profiles at the end of two successive slots. \revise{The transition probability $\P{\ellv' \to \ellv}$ is derived in Appendix~\ref{app:process_battery}.} 
 %    Let also $u_{(x',b'),(x,b)}$ be the number of devices whose process-battery state goes from $(x',b')$ to $(x,b)$. % in the transition from $\ellv'$ to $\ellv$. 
 %    For convenience, we also index $(x,b)$ by its position in %the sequence 
 %    $\{(0,0), \dots, (0,\Er), (1,0), \dots,(1,\Er)\}$. That is, we also write $\ellv'$ as $(\ell'_1, \dots, \ell'_{2\Er+2})$, $\ellv$ as $(\ell_1, \dots, \ell_{2\Er+2})$, and $u_{(x',b'),(x,b)}$ as $u_{j,k}$ where $j$ and $k$ are the associated indices of $(x',b')$ and $(x,b)$, respectively.
	% The transition probabilities of $\Lm$ %between the profiles of the $\Ur-1$ devices 
	% are %given by
	% \begin{align} \label{eq:trans_profile}
	% 	\P{\ellv' \to \ellv} 
 %        &= \sum_{\{u_{(x',b'),(x,b)}\}}  \bigg(\prod_{(x',b'),(x,b) \in \{0,1\} \times [0\;:\;\Er]} \notag \\
 %        &\qquad \qquad \qquad \qquad \P{(x',b') \to (x,b)}^{u_{(x',b'),(x,b)}}\bigg)  \notag\\
	% 	&\qquad \cdot \prod_{j=1}^{2\Er+2} \prod_{k=1}^{2\Er+2} \binom{\ell'_j}{u_{j,1}} \binom{\ell'_j - \sum_{q=1}^{k-1} u_{j,q}}{u_{j,k}}.
	% \end{align}
	% where the sum is over all values of $\{u_{(x',b'),(x,b)}\}$ such that 
 %    % \begin{align}
 %        $u_{(x',b'),(x,b)} \in [0:\min\{\ell'_{(x',b')},\ell_{(x,b)}\}]$, %\\ 
	%    $\ell'_{(x',b')} = \textstyle\sum_{(x'',b'')} u_{(x',b'),(x'',b'')}$,  %\\
 %        and $\ell_{(x,b)} = \textstyle\sum_{(x'',b'')} u_{(x'',b''),(x,b)}$. 
 %    % \end{align}
	The steady-state distribution of $\Lm$ is ${\rm Mul}(\Ur-1,2\Er+2,\{\nu_{(x,b)}\}_{x\in\{0,1\}, b\in[0:\Er]})$.
 
	% in Fig.~\ref{fig:markov_battery}.
	%From these transition probabilities, we compute the steady-state distribution $\mu_{\rm WED}_{\ellv}$, $\ellv \in \Lc$ of the battery profile by solving the balance equations.
	\subsection{$\!$Markov Chain Describing the Operation of a Generic Device} \label{sec:Markov_G}
	%We now define a Markov chain that fully describes the operation of a device. We first denote by $B\of{n} \in [0:\Er]$ the battery level of the device of interest at the end of slot $s$. Next, we denote the process-battery profile of the remaining $\Ur-1$ devices at the end of slot $s$ 
	%by $\Lm\of{n} \in \Lc$. Finally, recall that we denote the current state of the device and its estimate by the receiver as $X\of{n} \in \{0,1\}$ and  $\hat X\of{n} \in \{0,1\}$, respectively. 
	 The Markov chain %, whose state is defined as 
	$G\of{n} = (X\of{n},\hat X\of{n},B\of{n},\Lm\of{n})$ fully characterizes the operation of a generic device across slots. The transition probabilities of the chain~$G\of{n}$ are given by%. For $X\of{n-1} \in \{\refresh,\fail\}$, $X\of{n} = \refresh$ if, in slot $s$, the device successfully delivers a new update. It follows that
	\begin{multline} 
		\P{(x', \hat x', {b'},\ellv') \to (x, \hat x, b,\ellv)} \\ =  \P{\ellv' \to \ellv}\P{(x', \hat x', b') \to (x, \hat x, b) \;|\; \ellv'}. \label{eq:tmp422}
	\end{multline}
	Here,  %$\P{x, b'\!\to\! b \;|\; \ellv'}$ is a shorthand for $\P{X\of{n} \!=\! x, B\of{n} \!=\! b \;|\; B\of{n-1} \!=\! b', \Lm\of{n-1} \!=\! \ellv'}$, and 
	$\P{(x', \hat x', b') \to (x, \hat x, b) \;|\; \ellv'}$ is given by 
    \begin{align} \label{eq:PtransG_x=x'}
		&\P{(x', \hat x', b') \to (x, \hat x, b) \;|\; \ellv'} =\notag \\
		&\begin{cases}
			q_{x'x'}\big[ \pi_{b'}\of{x'x'} (1\!-\omega_{b',\ellv'})\phi\sups{trans.}_{b} +  (1\!-\pi_{b'}\of{x'x'})\phi\sups{no~trans.}_{b'\to b}\big], \\
			\qquad\qquad\qquad\qquad \qquad\quad~ \text{if~$\hat x \ne x' \ne \hat x'$}, \\
			 q_{x'x'}\big[ \pi_{b'}\of{x'x'} \phi\sups{trans.}_{b} +  (1-\pi_{b'}\of{x'x'})\phi\sups{no~trans.}_{b'\to b}\big], \\ \qquad \qquad\qquad\qquad\qquad\quad \text{if~$\hat x = x' = \hat x'$}, \\
			 q_{x'x'} \pi_{b'}\of{x'x'} \omega_{b',\ellv'}\phi\sups{trans.}_{b},
			 ~ \quad \text{if~$\hat x = x' \ne \hat x'$}, \\
			 0,  \qquad \qquad\qquad\qquad\qquad~ \text{if~$\hat x \ne x' = \hat x'$}.
		\end{cases} \!\!\!\!\!\!\!\!\!\!\!\!\!\!
	\end{align}
	if $x = x'$, and
	\begin{align} \label{eq:PtransG_x_ne_x'}
		&\P{(x', \hat x', b') \to (x, \hat x, b) \;|\; \ellv'} =\notag \\
		&\begin{cases}
			q_{x'x}\big[ \pi_{b'}\of{x'x} (1-\omega_{b',\ellv'}) \phi\sups{trans.}_{b} +  (1 -\pi_{b'}\of{x'x})\phi\sups{no~trans.}_{b'\to b}\big], \\
			\qquad\qquad\qquad\qquad \qquad\quad~ \text{if~$\hat x = x' = \hat x'$}, \\
			 q_{x'x} \pi_{b'}\of{x'x} \omega_{b',\ellv'}\phi\sups{trans.}_{b},\qquad \text{if~$\hat x \ne x' = \hat x'$}, \\
			 q_{x'x}\big[ \pi_{b'}\of{x'x} \phi\sups{trans.}_{b} +  (1-\pi_{b'}\of{x'x})\phi\sups{no~trans.}_{b'\to b}\big], \\
			 \qquad \qquad\qquad\qquad\qquad\quad~ \text{if~$\hat x \ne x' \ne \hat x'$}, \\
			 0,  \qquad \qquad\qquad\qquad\qquad~ \text{if~$\hat x = x' \ne \hat x'$}.
		\end{cases} \!\!\!\!\!\!\!\!\!\!\!\!\!\!
	\end{align}
    if $x\ne x'$. \revise{See Appendix~\ref{app:Gn} for the detailed derivation.}

	\section{Average Penalty Analysis} \label{sec:AoII}
	We now analyze the average penalty of a generic device. %We first provide some useful preliminaries results that will facilitate the derivation.
	% \subsection{\gls{AoII} Analysis via the Wrong/Correct-Estimate Duration}
	%As shown in Fig.~\ref{fig:AoI_process}, we denote by $Y$ the {\em inter-refresh time}, i.e., the number of slots that elapse between two successive \gls{AoI} refreshes for the device of interest. %This interval does not generally correspond to the time between two successful update deliveries. Indeed, the reception of an update triggers an \gls{AoI} refresh only if the update has not been available at the gateway. (In the chain $G\of{n}$ analyzed in Section~\ref{sec:Markov_G}, if $X\of{n-1} \in \{\refresh, \Sc\}$, then $X\of{n} = \refresh$ only if a new update is delivered; see~\eqref{eq:tmp422}.) We further denote by $Z$ the \gls{AoI} value right after a refresh, and refer to it as the \emph{refresh value}. 
	%After a refresh, the current \gls{AoI} is set to $1$. 
	%The discretized \gls{AoI} \gls{PMF}, peak \gls{AoI} \gls{PMF}, average \gls{AoI}, and \gls{AVP} can be expressed in terms of the distribution of $Y$ as follows.
    For convenience, we denote by $\{R\of{i}\}_i$ the sequence of time instants at which $\delta\of{n}$ changes from $1$ to $0$, and by $\{S\of{i}\}_i$ the time instants at which $\delta\of{n}$ changes from $0$ to $1$. 
    We denote by $W\of{i} = R\of{i} - S\of{i}$ the duration of the $i$th period over which the receiver has a wrong estimate, and by $Y\of{i} = S\of{i+1} - R\of{i}$ the duration of the $i$th period with a correct estimate. These quantities are also depicted in Fig.~\ref{fig:AoII_process}. We refer to the random variables $W$ and $Y$ whose realizations are determined by these durations as the \gls{WED} and \gls{CED}, respectively. Note that the process state does not change within a wrong/correct-estimate period. {We denote by $\tilde{X}\of{i}$ the process state during the $i$th wrong-estimate period and let $W_x\of{i}$ represent the conditional \gls{WED} process $\{W\of{i}\}_{i \colon \tilde{X}\of{i} = x}$, $x \in \{0,1\}$. Let $\tilde{X}$ and $W_x$ be random variables whose distributions are the stationary distributions of $\tilde{X}\of{i}$ and $W_x\of{i}$, respectively.
    }
    %{let random variable $\tilde{X}$ take values from the process $\{\tilde{X}\of{i}\}_i$.} Let {random variable} $W_{x}$ {take values from} the conditional \gls{WED} process $\{W\of{i} \colon \tilde{X}\of{i} = x\}_{{i}}$ for $x \in \{0,1\}$. %The conditional \gls{CED} $Y_{x}$ is defined similarly. %We also use the average error duration $\E{W}$ as a performance indicator. 
    The average penalty can be computed in terms of these random variables %\gls{WED} and \gls{CED} 
    as follows. 
	\begin{theorem}[Average penalty] \label{th:avg_penalty}
        For a penalty function~$f$, the average penalty defined in~\eqref{eq:avg_penalty} is given by
        % \begin{equation}
            $\overline{F} = \frac{\Gamma}{\E{W} + \E{Y}}$ %\label{eq:avg_penalty_WY}
        % \end{equation}
        with {$\Gamma = \sum_{x = 0}^1 \mathbb{P} \big[\tilde X = x\big] \EE%_{\tilde{X}, W_{\tilde{X}}} 
            \big[\sum_{j=1}^{W_{x}} f_{x}(j)\big]$.} %, {where the expectation is first over $W_{\tilde{X}}$ for a given value of $\tilde{X}$ and then over $\tilde{X}$.}
        In particular, {for the power penalty function $f_{X\of{n}} =  (\Lambda\of{n})^{\alpha_{X\of{n}}}$, if $\alpha_0$ and $\alpha_1$ are nonnegative integers, we have that
        \begin{equation}
            \!\Gamma = \sum_{x = 0}^1 \P{\tilde X \!=\! x} \frac{1}{\alpha_{x} \!+\! 1} \sum_{k=0}^{\alpha_{x}} \binom{\alpha_{x} \!+\! 1}{k} \Br_{k} \E{W_{x}^{\alpha_{x}-k + 1}} \label{eq:pen_power_WY}
        \end{equation}
        where $\{\Br_{k}\}$ are the Bernoulli numbers.}
        The average \gls{AoII}
        % and \gls{AoII} violation probability are
        is given by $\overline{F}$ with $\Gamma = \frac12 \big(\E{W} + \E{W^2}\big)$.
        
	\end{theorem}
	\begin{proof}
        \revise{See Appendix~\ref{proof:avg_penalty}.}
	\end{proof}
    
    Theorem~\ref{th:avg_penalty} implies that to compute the average penalty, we need to know the distribution of %\gls{WED}-wise process state 
    $\tilde{X}$, %the conditional \gls{WED} 
    $W_{x}$, and %the \gls{CED} 
    $Y$.
	%	\begin{remark} \label{rem:other_metric}
		%		While we focus on the average \gls{AoI} and \gls{AVP}, we note that Many other \gls{AoI} metrics can be derived from the distribution of $Y$. For example, the average peak \gls{AoI} is given by $1 + \E{Y}$. We choose 
		%	\end{remark}
	% \subsection{Average AoII} \label{sec:avg_AoI}
	One can derive these distributions using a first-step Markovian analysis similar to~\cite[Sec.~IV-B]{Ngo24statusupdate}. However, this requires computing the transition probabilities between all $4(\Er + 1)\binom{\Ur+2\Er+1}{2\Er+2}$ states of $G\of{n}$, which is cumbersome for large $\Ur$ and $\Er$. %In the next section, 
    We next propose an approximation that does not require this computation.
		
		%-----------------------------------------------
		\section{Proposed Approximation} \label{sec:AoII_approx}
		To avoid the complexity issue just described, {similar to~\cite{Ngo24statusupdate},} we ignore the time dependency of the process-battery profile of the other devices. Specifically, we assume the following.
		\begin{simplification} \label{simplification}
			Given a device of interest, the process-battery profile $\Lm$ of the remaining $\Ur-1$ devices evolves according to a stationary memoryless process across slots. 
		\end{simplification}
		%This simplification allows us to analyze the behavior of the system for large $\Ur$ and $\Er$ values, and, as we shall see, results in tight approximations of the average AoI and AVP for all scenarios explored in Section~\ref{sec:results}. 
        
		Under this simplification, the successful-decoding probability of an update transmitted with $b$ energy units is given by %the average of $\omega_{b,\Lm}$ over $\Lm$, i.e., %by $\bar{\omega}_b$ given in~\eqref{eq:avg_w}.
        %At steady state, the average successful decoding probability of an update transmitted with $b$ energy units is
	% \begin{equation} \label{eq:avg_w}
		$\bar{\omega}_b = \E{\omega_{b,\Lm}}$,
	% \end{equation}
	where the expectation is over the steady-state distribution %$ follows the multinomial distribution with number of trials $\Ur-1$, number of events $B+1$, and event probabilities $\{\nu_i\}_{i = 0}^\Er$, denoted by $
	%${\rm Mult}(\Ur-1,2\Er+2,\{\nu_{(x,b)}\}_{x \in\{0,1\}, b \in [0:\Er]})$ 
    of $\Lm$. %Here, $\{\nu_{(x,b)}\}_{x\in\{0,1\}, b\in[0:\Er]}$, computed in Section~\ref{sec:source_battery_evolution}, is the steady-state distribution of the process-battery state. 
		% \begin{align}
			%     \bar{\omega}_b = \E[\Lm]{\omega_{b,\Lm}}.
			% \end{align}
		%This allows us to simplify the Markov chain as follows. %and derive the distribution of $\tilde{X}$, $W$, and $Y$ in closed form, as presented next. %, allowing us to compute the average \gls{AoI} and age-violation probability.
		% \subsection{Simplified Markov Chains}
		%Under Simplification~\ref{simplification}, 
        The process-battery profile $\Lm$ in each slot is drawn independently from this distribution. % ${\rm Mul}(\Ur-1,2\Er+2,\{\nu_{(x,v)}\}_{x \in \{0,1\}, b \in [0:\Er]}$. 
        %We therefore only need to track the process state $X\of{n}$, its estimate $\hat X \of{n}$, and the battery level $B\of{n}$. % of the associated device. 
       Therefore, the Markov chain describing the operation of the device is reduced to $(X\of{n},\hat X\of{n}, B\of{n})$, obtained by grouping the states of  $G\of{n}$ as follows. We partition the state space of $G\of{n}$ into disjoint subsets of the form $\{(x,\hat x,b,\ellv) \colon \ellv \in \Lc\}$, and represent each subset by a state $(x,\hat x,b)$. We then compute the transition probabilities between states of $(X\of{n}, \hat X\of{n} B\of{n})$ as $\P{(x',\hat x',b') \to (x,\hat x,b)} = \E{\sum_{\ellv \in \Lc} \P{(x',\hat x',b',\Lm') \to (x,\hat x,b,\ellv)}}$, where the expectation is over the steady-state distribution of $\Lm'$.
		 Specifically, if $x = x'$, we have that
		\begin{align}
			&\P{(x', \hat x', b') \to (x', \hat x, b)} = \notag \\
			&\begin{cases}
				q_{x'x'}\big[ \pi_{b'}\of{x'x'} (1-\bar{\omega}_{b'})\phi\sups{trans.}_{b} +  (1-\pi_{b'}\of{x'x'})\phi\sups{no~trans.}_{b'\to b}\big], \\
				\qquad\qquad\qquad\qquad \qquad~ \text{if~$\hat x \ne x' \ne \hat x'$}, \\
				q_{x'x'}\big[ \pi_{b'}\of{x'x'} \phi\sups{trans.}_{b} +  (1-\pi_{b'}\of{x'x'})\phi\sups{no~trans.}_{b'\to b}\big], \\
				\qquad \qquad \qquad \qquad \qquad~ \text{if~$\hat x = x' = \hat x'$}, \\
				q_{x'x'} \pi_{b'}\of{x'x'} \bar{\omega}_{b'}\phi\sups{trans.}_{b},\quad~ \text{if~$\hat x = x' \ne \hat x'$}, \\
				0, \qquad \qquad\qquad\qquad\quad~ \text{if~$\hat x \ne x' = \hat x'$}.
			\end{cases} \!\!\!\!\!\!\!\!\!\!\!\!\!   \label{eq:ptrans_simp_chain_x_=_x'}
		\end{align}
		If $x \ne x'$, %for $x' \in \{\refresh,\fail\}$, 
		we have that
		\begin{align}
			&\P{(x', \hat x', b') \to (x, \hat x, b)} = \notag \\
			&\begin{cases}
				q_{x'x}\big[ \pi_{b'}\of{x'x} (1-\bar{\omega}_{b'}) \phi\sups{trans.}_{b} +  (1-\pi_{b'}\of{x'x})\phi\sups{no~trans.}_{b'\to b}\big], \\
				\qquad\qquad\qquad\qquad \quad~ \text{if~$\hat x = x' = \hat x'$}, \\
				q_{x'x} \pi_{b'} \bar{\omega}_{b'}\phi\sups{trans.}_{b},\qquad \text{if~$\hat x \ne x' = \hat x'$}, \\
				q_{x'x}\big[ \pi_{b'}\of{x'x}\phi\sups{trans.}_{b} +  (1-\pi_{b'}\of{x'x})\phi\sups{no~trans.}_{b'\to b}\big], \\
				\qquad \qquad\qquad\qquad\quad~ \text{if~$\hat x \ne x' \ne \hat x'$}, \\
				 0,  \qquad \qquad\qquad\qquad~ \text{if~$\hat x = x' \ne \hat x'$}.
			\end{cases} \!\!\!\!\!\!\!\!\label{eq:ptrans_simp_chain_x_ne_x'}
		\end{align}

        From the derived transition probabilities of the chain $(X\of{n}, \hat X\of{n}, B\of{n})$, we computed the steady state distribution $\{p_{(x,\hat x,b)}\}$ by solving the balance equations. \revise{By analyzing this chain, we can derive in closed form the %distribution of $W$, $Y$, and the related quantities necessary for the computation of the
        average penalty and the \gls{MEP} under Simplification~\ref{simplification}. We show next how to compute $\E{W}$ and $\Gamma$. The derivation of $\E{Y}$ and the \gls{MEP} can be found in Appendices~\ref{app:approx_CED} and~\ref{app:approx_MEP}, respectively.}
        
		%\subsection{Approximate \gls{WED} Distribution and Related Quantities}
        We modify the chain $(X\of{n}, X\of{n}, B\of{n})$ to obtain a Markov chain that describes the slot-wise evolution of a device within a wrong-estimate period. Specifically, we combine all states $(x,\hat x,b)$ with $x = \hat x$ and $b \in [0:\Er]$ into a single state $\Ar\sub{CE}$ that represents all slots with a correct estimate. % that represents an \gls{AoI} refresh. In other words, we redirect all transitions that lead to an \gls{AoI} refresh into a new state $\refresh''$.    
		%Furthermore, we split the state $b \in [\Er]$ in $M_1$ into two states $\Sc_b$ and $\fail,0$, specifying whether the latest update of the device has or has not reached the gateway, respectively, before the current slot. 
		%Note that if the device is in state $\Sc_i$, $i\in [\Er]$, a retransmission of the latest update cannot trigger an \gls{AoI} refresh. 
		We refer to the resulting Markov chain as $M\sub{WE}$ and depict it in Fig.~\ref{fig:markov_WED}. 
		This chain is a terminating Markov chain (see~\cite[App.~B]{Ngo24statusupdate}) with an absorbing state $A\sub{CE}$ and $2\Er + 2$ transient states $\{(x,\hat x,b)\}_{x \ne \hat x \in \{0,1\}, b \in [0:\Er]}$. We denote the transition probability matrix of this chain as $\bigg[\begin{matrix}
			\Ts\sub{WE} ~~ \av\sub{WE} \\ \mathbf{0} ~~ 1
		\end{matrix}\bigg]$, where $\Ts\sub{WE}$ contains the probabilities of transitions between the transient states and $\av\sub{WE}$ contains the probabilities of transitions from the transient states to the absorbing state. We obtain $\Ts\sub{WE}$ and $\av\sub{WE}$ using~\eqref{eq:ptrans_simp_chain_x_=_x'} and~\eqref{eq:ptrans_simp_chain_x_ne_x'}.

         	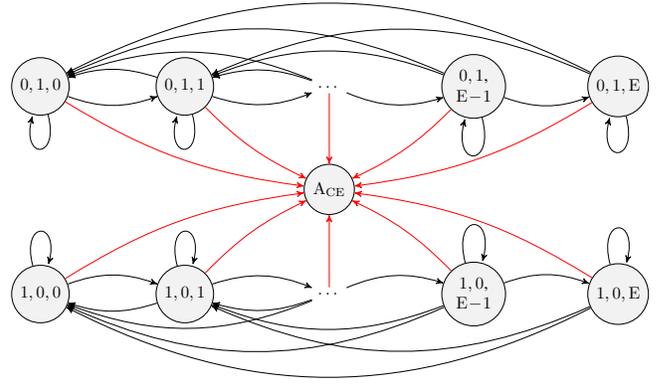
\begin{figure}[t!]
			\centering
						\vspace{-.3cm}
			\scalebox{.64}{\begin{tikzpicture}
					\tikzset{node distance=3cm, % Minimum distance between two nodes. Change if necessary.
						every state/.style={ % Sets the properties for each state
							semithick,
							fill=gray!10},
						initial text={},     % No label on start arrow
						double distance=4pt, % Adjust appearance of accept states
						every edge/.style={  % Sets the properties for each transition
							draw,
							->,>=stealth',     % Makes edges directed with bold arrowheads
							auto,
							semithick}}
					\node[state] (b0) {$1,0,0$};
					\node[state, right of= b0] (b1) {$1,0,1$};
					\node[right of= b1] (mid) {$\dots$};
					\node[state, right of= mid,align=center] (b3) {$1,0$, \\ $\Er\!-\!1$};
					\node[state, right of= b3] (b4) {$1,0,\Er$};
					
					\node[state, above=1.5cm of mid] (A) {$\Ar\sub{CE}$};
					
					\draw (b0) edge[loop above] node[above,midway] {} (b0);
					\draw (b1) edge[loop above] node[above,midway] {} (b1);
					\draw (b3) edge[loop above] node[above,midway] {} (b3);
					\draw (b4) edge[loop above] node[above,midway] {} (b4);
					
					\draw (b0) edge[bend left=20] node[above,midway] {} (b1);
					\draw (b1) edge[bend left=20] node[above,midway] {} (mid);
					\draw (mid) edge[bend left=20] node[above,midway] {} (b3);
					\draw (b3) edge[bend left=20] node[above,pos=.5] {} (b4);
					
					\draw (b1) edge[bend left=18] node[above=-.07cm,midway] {} (b0);
					\draw (mid) edge[bend left=21] node[above,midway] {} (b0);
					\draw (b3) edge[bend left=24] node[above=-.07cm,pos=.54,rotate=-2] {} (b0);
					\draw (b4) edge[bend left=27] node[above=-.05cm,pos=.5] {} (b0);
					
					%				\draw (b1) edge[bend left=15] node[above=-.07cm,midway] {$\rho_1$} (b1);
					\draw (mid) edge[bend left=18] node[above,midway] {} (b1);
					\draw (b3) edge[bend left=21] node[above=-.07cm,pos=.5] {} (b1);
					\draw (b4) edge[bend left=24] node[above=-.07cm,pos=.5] {} (b1);
					
					\draw[color=red] (b0) edge[bend left=12] node[above=-.07cm,midway] {} (A);
					\draw[color=red] (b1) edge[bend left=10] node[above=-.07cm,midway] {} (A);
					\draw[color=red] (mid) edge node[above,midway] {} (A);
					\draw[color=red] (b3) edge[bend right=10] node[above=-.07cm,pos=.54,rotate=-2] {} (A);
					\draw[color=red] (b4) edge[bend right=12] node[above=-.05cm,pos=.5] {} (A);

                    %%%%%%%%%%%%%%%
					\node[state, above=3.1cm of b0] (b10) {$0,1,0$};
					\node[state, right of= b10] (b11) {$0,1,1$};
					\node[right of= b11] (mid1) {$\dots$};
					\node[state, right of= mid1,align=center] (b13) {$0,1$, \\ $\Er\!-\!1$};
					\node[state, right of= b13] (b14) {$0,1,\Er$};
					
					% \node[state, below of= b0] (A) {$\Ar_{\setminus(x,\hat x)}$};

					\draw (b10) edge[loop below] node[above,midway] {} (b10);
					\draw (b11) edge[loop below] node[above,midway] {} (b11);
					\draw (b13) edge[loop below] node[above,midway] {} (b13);
					\draw (b14) edge[loop below] node[above,midway] {} (b14);
					
					\draw (b10) edge[bend right=20] node[above,midway] {} (b11);
					\draw (b11) edge[bend right=20] node[above,midway] {} (mid1);
					\draw (mid1) edge[bend right=20] node[above,midway] {} (b13);
					\draw (b13) edge[bend right=20] node[above,pos=.5] {} (b14);
					
					\draw (b11) edge[bend right=18] node[above=-.07cm,midway] {} (b10);
					\draw (mid1) edge[bend right=21] node[above,midway] {} (b10);
					\draw (b13) edge[bend right=24] node[above=-.07cm,pos=.54,rotate=-2] {} (b10);
					\draw (b14) edge[bend right=27] node[above=-.05cm,pos=.5] {} (b10);
					
					%				\draw (b1) edge[bend left=15] node[above=-.07cm,midway] {$\rho_1$} (b1);
					\draw (mid1) edge[bend right=18] node[above,midway] {} (b11);
					\draw (b13) edge[bend right=21] node[above=-.07cm,pos=.5] {} (b11);
					\draw (b14) edge[bend right=24] node[above=-.07cm,pos=.5] {} (b11);

                    \draw[color=red] (b10) edge[bend right=12] node[above=-.07cm,midway] {} (A);
					\draw[color=red] (b11) edge[bend right=10] node[above=-.07cm,midway] {} (A);
					\draw[color=red] (mid1) edge node[above,midway] {} (A);
					\draw[color=red] (b13) edge[bend left=10] node[above=-.07cm,pos=.54,rotate=-2] {} (A);
					\draw[color=red] (b14) edge[bend left=12] node[above=-.05cm,pos=.5] {} (A);
				\end{tikzpicture}
			}

            \vspace{-.3cm}
			\caption{Markov chain $M\sub{WE}$ describing the slot-wise evolution of a device within a wrong-estimate period, i.e., when $(X,\hat X) \in \{(0,1),(1,0)\}$. The absorbing state $\Ar\sub{CE}$ %indicates the first slot where the receiver's estimate becomes correct. It 
            represents all states of the chain $(X\of{n},\hat X\of{n}, B\of{n})$ where $X\of{n} = \hat X\of{n}$.}
			\label{fig:markov_WED}
						\vspace{-.5cm}
		\end{figure}
  
        We observe that the device experiences a wrong-estimate period when the chain $(X\of{n}, \hat X\of{n}, B\of{n})$ enters one of the transient states of $M\sub{WE}$. %, i.e., when $(X\of{n}, \hat X\of{n}, B\of{n})$ becomes similar to one of the transient states of $M_{01}$ or $M_{10
		Let $\P{\to (x,\hat x,b)}$ be the probability that the state $(x,\hat x,b)$  is visited after a state with a different value of $(X, \hat X)$. We can expressed this probability as
			\begin{multline}
				\P{\to (x,\hat x,b)} = \\ \sum_{(x', \hat x') \ne (x,\hat x), b' \in [0:\Er]} p_{(x,\hat x,b)} \P{(x', \hat x',b') \to (x,\hat x, b)}.
			\end{multline}
		%where $p_{(x,\hat x,b)}$ is the steady-state probability of the state  $(x,\hat x,b)$ of the chain $(X\of{n}, \hat X\of{n}, B\of{n})$, computed from the transition probabilities. 
        Finally, we obtain the probability that a wrong-estimation period starts from state $(x,\hat x,b)$ from the normalization 
        \begin{equation}
            \tau_{(x,\hat x,b)} = \frac{\P{\to (x,\hat x,b)}}{\sum_{(x',\hat x') \in \{(0,1),(1,0)\}, b' \in [0:\Er]} \P{\to (x',\hat x',b')}}
        \end{equation}
        for $x \ne \hat x$.
        We denote $\tauv\sub{WE} = (\tau_{(0,1,0)}, \dots, \tau_{(0,1,\Er)},\tau_{(1,0,0)}, \dots, \tau_{(1,0,\Er)})$. 
		
		The \gls{WED} $W$ corresponds to the absorption time of the chain $M\sub{WE}$ when starting with initial probability vector $\tauv\sub{WE}$. Therefore, $W$ follows the discrete phase-type distribution characterized in~\cite[Sec.~2.2]{Neuts1994} (see also~\cite[Lem.~3]{Ngo24statusupdate}). The \gls{PMF} and moments of $W$ are %given by
		\begin{align} 
			\!\!\P{W = w} &= \tauv\sub{WE}^\T \Ts\sub{WE}^{w-1}\av\sub{WE}, \quad w = 1,2,\dots \label{eq:approx_dist_W} \\
			\E{W} &=  \tauv\sub{WE}^\T (\Is-\Ts\sub{WE})^{-1} \mathbf{1}, \label{eq:approx_EW} \\
			\E{W^2} &= 2\tauv\sub{WE}^\T (\Is-\Ts\sub{WE})^{-2}  \mathbf{1} - \E{W}. \label{eq:approx_EW2}
		\end{align}
        This allows us to compute $\Gamma$ for the average \gls{AoII}.
        
        Let $W_{(x,\hat x, b)}$ denote the conditional \gls{WED} given that the device enters the wrong-estimate period via the state $(x,\hat x, b)$ of the chain $M\sub{WE}$. Under this condition, we have that 
        \begin{align}
            \E{{W_{(x,\hat x, b)}}} &= \ev^\T_{(x,\hat x, b)} (\Is-\Ts\sub{WE})^{-1} \mathbf{1}, \label{eq:E_Wx} \\
            \E{{W^2_{(x,\hat x, b)}}} &=  2\ev^\T_{(x,\hat x, b)} (\Is-\Ts\sub{WE})^{-2} \mathbf{1} - \E{W_{(x,\hat x, b)}}, \label{eq:E_Wx2}   
        \end{align}
        where $\ev_{(x,\hat x, b)}$ is the one-hot vector indicating the position of $(x,\hat x, b)$ in $\{(0,1,0), \dots, (0,1,\Er),(1,0,0), \dots, (1,0,\Er)\}$. 
        % It follows that
        % \begin{align}
        %     \E{\alpha_{\tilde{X}} W_{\tilde{X}}} &= \sum_{x,\hat x,b} \tau_{(x,\hat x, b)} \alpha_x \ev^\T_{(x,\hat x, b)} (\Is-\Ts\sub{WE})^{-1} \mathbf{1}, \\
        %     \E{\alpha_{\tilde{X}} W^2_{\tilde{X}}} &= 2\sum_{x,\hat x,b} \tau_{(x,\hat x, b)} \alpha_x \ev^\T_{(x,\hat x, b)} (\Is -\!\Ts\sub{WE})^{-2} \mathbf{1} \notag \\
        %     &\quad - \E{\alpha_{\tilde{X}} W_{\tilde{X}}},
        % \end{align}
        % where the sums are over $(x,\hat x) \in \{(0,1),(1,0)\}$, $b \in [0:\Er]$.
        This allows us to compute $\Gamma$  for the linear penalty function as
        \begin{equation}
            \Gamma = \!\!\sum_{(x,\hat x) \in \{(0,1),(1,0)\}, b \in [0:\Er]} \!\!\!\!\!\!\!\!\tau_{(x,\hat x, b)} \alpha_x \ev^\T_{(x,\hat x, b)} (\Is -\Ts\sub{WE})^{-2} \mathbf{1}.
        \end{equation}

				\section{Numerical Experiments} \label{sec:results}
                
				In this section, we assume a specific slot-wise channel model and apply the analytical results to numerically evaluate the {penalty function and the \gls{MEP}}. %We first describe the considered channel model in details and derive the successful decoding probability~$\omega_{b,\ellv}$.
				%\subsection{Channel Model and Successful Delivery Probability} \label{sec:succ_prob}
				Specifically, we assume that a slot comprises $\Nr$ uses of a real-valued \gls{AWGN} channel. %, {which is relevant in systems where the devices estimate their channel based on downlink pilot broadcast from the gateway and pre-equalize their uplink signal}. 
                In a slot, active device~$i$ with battery level $b_i$ transmits a signal $\sqrt{b_{i}/\Nr}\Xm_i \in \RR^\Nr$ with $\|\Xm_i\| = 1$. The received signal is 
				$
				\Ym = \sum_{i\in \Uc\sub{a}} \sqrt{b_{i}/\Nr}  \Xm_i + \Zm,
				$
				where $\Uc\sub{a}$ is the set of active devices and
				%where $K$ is the number of active devices %, $\Xm_i \in \RR^\Nr$ (with $\|\Xm_i\| = 1$) is the transmitted signal of device~$i$, $b_i$ is the battery level of device~$i$, 
				$\Zm \sim \Nc(\mathbf{0},\sigma^2\Is)$ is the \gls{AWGN}. %Let $b_i$ be the battery level of device $i$, we have that $P_i = b_i/\Nr$. 
				The devices transmit at rate $\Rr$ bits/channel use, i.e., $\Xm_i$ belongs to a codebook containing $2^{\Nr\Rr}$ codewords. We %consider shell codes for which 
                let the codeword be uniformly distributed on the unit sphere. 
                %We analyze two decoding scenarios: i) decoding without capture, where decoding is attempted only on packets transmitted in singleton slots; ii) decoding with capture, where the receiver attempts to decode every packet transmitted in a slot using \gls{SIC}. In both cases, we derive the successful decoding probability~$\omega_{b,\ellv}$ in a similar manner as in~\cite{Ngo24statusupdate}. 
				% \subsubsection{Without capture} 
				% In this scenario, 
                We assume that all collided packets are lost, i.e.,  decoding is attempted only on packets transmitted in singleton slots. %This model allows us to revisit the collision channel model commonly used in modern random-access analyses, e.g.~\cite{Liva2011,Munari2020modern,Demirhan2019}, and further account for single-user decoding errors due to finite-blocklength effects. 
                We derive the successful decoding probability~$\omega_{b,\ellv}$, accounting for single-user decoding errors due to finite-blocklength effects, in a similar manner as in~\cite[Sec.~VII-A-1]{Ngo24statusupdate}. 
        Hereafter, we consider $\Ur=1000$ devices with battery capacity $\Er = 8$, a slot length~$\Nr$ of $100$ channel uses, transmission rate $\Rr$ of $0.8$ bits/channel use, and noise variance $\sigma^2 = -20$~dB. 
        %We also consider $\Ur=1000$ %sources with identical probability of state transition $\bar{q} = q_{10} = q_{01}$. The battery capacity and energy harvesting rate are 
        %devices with battery capacity $\Er = 8$. % and $\gamma = 0.005$. 
        %We focus on the case without capture. 
        For convenience, we denote the average probability for a process to change state as $\bar{q} = 2 q_{01} q_{10}/(q_{01} + q_{10})$. {We examine the reactive, random, and hybrid strategies described in Section~\ref{sec:protocol}.}
        \revise{For each strategy, we numerically optimize the transmission probabilities $\Pim$ to minimize the average penalty or average \gls{AoII} using the Nelder-Mead simplex algorithm %~\cite{nelder1965simplex} 
        with multiple initializations. %, \hoang{a commonly-used search method for multidimensional nonlinear optimization. However, we note that this heuristic method can converge to nonstationary points and is highly sensitive to the initial values. To circumvent this issue, we try multiple initializations and run the optimization multiple times.}
        }
        %As in~\cite{Munari24_AoII}, we examine three strategies: i) a \emph{reactive} strategy, where the device only transmits when there is a process state change, i.e., $\pi_b\of{ij} = 0$ if $i = j$; ii) a \emph{random} strategy, where the device uses the same transmission probability regardless of its process state, i.e., $\pi_b\of{ij} = \pi_b$ for every $(i,j)$; iii) a \emph{hybrid} strategy, where transmission probability $\pi_b\of{ij}$ can be chosen between $0$ and $1$ for every $(i,j,b)$. % \pp{PP: This should be elaborated upon more, although it was already proposed in [12]. Perhaps the last part of the previous section can discuss these strategies and relate them to the model developed previously in the paper - which is different from the one in [12].}
        
        \subsection{Symmetric Processes} \label{sec:avgAoII_minimization}
        
        We first consider symmetric processes, {i.e., $q_{10} = q_{01}$,} with energy harvesting rate $\gamma_0 = \gamma_1 = 0.005$. 
        In Fig.~\ref{fig:symmetric}(a), we show the average \gls{AoII} achieved by the three strategies with optimized transmission probabilities as functions of the average total number of transitions in a slot, $\Ur \bar{q}$. We use the approximation in Section~\ref{sec:AoII_approx} to evaluate the average \gls{AoII}, and also show the simulation result for the hybrid strategy, obtained from an implementation of the protocol over~$10^6$ slots. We observe that the simulation results coincide with our approximation, confirming the tightness of our proposed approximations. The reactive strategy achieves the highest average \gls{AoII} because if a transition  to a state different from the receiver's estimate is not reported, the receiver's estimate remains in error for a long period. The random strategy achieves a significantly lower average \gls{AoII} due to the ability to perform multiple attempts to report a state change. The hybrid strategy brings a small improvement upon the random strategy. 
        \begin{figure}[t!]
            \centering
            \subcaptionbox{Average \gls{AoII}}{
            \begin{tikzpicture}[scale=.9]
		\begin{axis}[%
			width=3.1in,
			height=1in,	
			scale only axis,
			unbounded coords=jump,
			xmin=0.001,
			xmax=1,
            xmode=log,
			xlabel={$\Ur \bar{q}$ (transitions/slot)},
            xlabel style={yshift=.1cm},
			ymode=log,
			ymin=10,
			ymax=2000,
			yminorticks=true,
			ytick= {1e1, 1e2, 1e3, 2e3},
            yticklabels = {$10^1$,$10^2$,$10^3$,$2\!\times\!10^3$},
			ylabel style={yshift=-.6cm,xshift=-.3cm},
			ylabel={average AoII, $\overline{\Lambda}$},
			axis background/.style={fill=white},
			title style={font=\bfseries},
			xmajorgrids,
            xminorgrids,
			ymajorgrids,
			yminorgrids,
			legend style={at={(.99,0.01)}, anchor=south east, legend cell align=left, align=left,draw=none, fill=white, fill opacity=.8,text opacity = 1}
			]

            \addplot [line width = 1.5,dashdotted,color=OliveGreen]
			table[row sep=crcr]{%
            1.0e-03 1.5047e+03 \\ 
2.5e-03 1.4989e+03 \\ 
5.0e-03 1.4894e+03 \\ 
7.5e-03 1.4801e+03 \\ 
				1.0e-02 1.4708e+03 \\ 
2.0e-02 1.4349e+03 \\ 
4.0e-02 1.3679e+03 \\ 
6.0e-02 1.3067e+03 \\ 
8.0e-02 1.2506e+03 \\ 
1.0e-01 1.1991e+03 \\ 
2.0e-01 9.9276e+02 \\ 
4.0e-01 7.3528e+02 \\ 
6.0e-01 5.8144e+02 \\ 
8.0e-01 4.7838e+02 \\ 
1.0e+00 4.0467e+02 \\ 
			};
            \addlegendentry{reactive};
            
			\addplot [line width = 1.5,dashed,color=blue]
			table[row sep=crcr]{%
            1.0e-03 4.8894e+00 \\ 
2.5e-03 1.2083e+01 \\ 
5.0e-03 2.3780e+01 \\ 
7.5e-03 3.5289e+01 \\ 
				1.0e-02 4.6064e+01 \\ 
2.0e-02 8.6591e+01 \\ 
4.0e-02 1.5391e+02 \\ 
6.0e-02 2.0663e+02 \\ 
8.0e-02 2.4814e+02 \\ 
1.0e-01 2.8092e+02 \\ 
2.0e-01 3.6553e+02 \\ 
4.0e-01 3.8238e+02 \\ 
6.0e-01 3.5276e+02 \\ 
8.0e-01 3.1848e+02 \\ 
1.0e+00 2.8739e+02 \\ 
			}
            %node [pos=.2,pin={[pin edge={solid,OliveGreen,<-}]90:without capture},inner sep=1pt] {}
            ;
            \addlegendentry{random};

            \addplot [line width = 1.5,color=red]
			table[row sep=crcr]{%
            1.0e-03 4.2361e+00 \\ 
2.5e-03 1.0485e+01 \\ 
5.0e-03 2.0734e+01 \\ 
7.5e-03 3.0683e+01 \\ 
				1.0e-02 4.0622e+01 \\ 
2.0e-02 7.7241e+01 \\ 
4.0e-02 1.4036e+02 \\ 
6.0e-02 1.9218e+02 \\ 
8.0e-02 2.3376e+02 \\ 
1.0e-01 2.6797e+02 \\ 
2.0e-01 3.6339e+02 \\ 
4.0e-01 3.8237e+02 \\ 
6.0e-01 3.5275e+02 \\ 
8.0e-01 3.1847e+02 \\ 
1.0e+00 2.8737e+02 \\ 
			};
            \addlegendentry{hybrid};

            \addplot [mark = x,only marks,mark size = 4,mark color = red]
			table[row sep=crcr]{%
                % 0.01 45.9911 \\
                % 0.04 154.9866 \\
                % 0.1 280.2045 \\
                % 0.4 383.1440 \\
                1 288.2212\\
                .75 325.6479 \\
                .5 368.1714 \\
                .25 381.6228 \\
                .1 264.3130 \\
                .075 219.7971 \\
                .05 168.4473 \\
                .025 93.6691 \\
                .01 39.8314 \\
                .0075 30.6132 \\
                .005 20.8205 \\
                .0025 10.6636 \\
			};
            
   %          \addplot [mark = o,mark color = blue,only marks,mark size = 3]
			% table[row sep=crcr]{%
   %              0.01 45.9911 \\
   %              0.04 154.9866 \\
   %              0.1 280.2045 \\
   %              0.4 383.1440 \\
   %              1 286.6724\\
			% };
            % \addlegendentry{random strategy, simulation};

   %          \addplot [line width = 1.5,color=blue,forget plot]
			% table[row sep=crcr]{%
			% 	0.01 32.8182\\
   %              0.02 61.4184 \\
   %              0.04 110.7188 \\
   %              0.06 165.8437 \\
   %              0.08 202.7469 \\
   %              0.1 212.0990 \\
   %              0.2 292.1408 \\ 
   %              0.4 321.4237 \\
   %              0.6 309.2381 \\
   %              .8 288.6051 \\
   %              1 263.8766 \\
			% }
   %          node [pos=.3,pin={[pin edge={solid,OliveGreen,<-}]-90:with capture},inner sep=1pt] {};

   %          \addplot [line width = 1.5,dashed,color=red,forget plot]
			% table[row sep=crcr]{%
			% 	0.01 32.2853\\
   %              0.02 61.0747 \\
   %              0.04 110.7188 \\
   %              0.06 156.8622 \\
   %              0.08 201.2042 \\
   %              0.1 211.3085 \\
   %              0.2 289.1327 \\ 
   %              0.4 321.1538 \\
   %              0.6 308.0796 \\
   %              .8 283.8101 \\
   %              1 261.9046 \\
			% };
        \end{axis}
        \end{tikzpicture}}
            \subcaptionbox{\gls{MEP} achieved with the strategies in Fig.~\ref{fig:symmetric}(a)}{\begin{tikzpicture}[scale=.9]
		\begin{axis}[%
			width=3.1in,
			height=1in,	
			scale only axis,
			unbounded coords=jump,
			xmin=0.001,
			xmax=1,
            xmode=log,
			xlabel={$\Ur \bar{q}$ (transitions/slot)},
            xlabel style={yshift=.1cm},
			ymode=log,
			ymin=1e-3,
			ymax=1,
			yminorticks=true,
			ylabel style={yshift=-.1cm},
			ylabel={\gls{MEP}},
			axis background/.style={fill=white},
			title style={font=\bfseries},
			xmajorgrids,
            xminorgrids,
			ymajorgrids,
			yminorgrids,
			legend style={at={(.99,0.01)}, anchor=south east, legend cell align=left, align=left,draw=none, fill=white, fill opacity=.8,text opacity = 1}
			]

            \addplot [line width = 1.5,dashdotted,color=OliveGreen]
			table[row sep=crcr]{%
            1.0e-03 1.5070e-03 \\ 
2.5e-03 3.7614e-03 \\ 
5.0e-03 7.5031e-03 \\ 
7.5e-03 1.1225e-02 \\ 
				1.0e-02 1.4928e-02 \\ 
2.0e-02 2.9545e-02 \\ 
4.0e-02 5.7879e-02 \\ 
6.0e-02 8.5063e-02 \\ 
8.0e-02 1.1115e-01 \\ 
1.0e-01 1.3620e-01 \\ 
2.0e-01 2.4753e-01 \\ 
4.0e-01 4.1587e-01 \\ 
6.0e-01 5.3443e-01 \\ 
8.0e-01 6.1518e-01 \\ 
1.0e+00 6.7435e-01 \\ 
			};
            \addlegendentry{reactive};
            
			\addplot [line width = 1.5,dashed,color=blue]
			table[row sep=crcr]{%
            1.0e-03 2.2454e-03 \\ 
2.5e-03 5.5937e-03 \\ 
5.0e-03 1.1129e-02 \\ 
7.5e-03 1.6613e-02 \\ 
				1.0e-02 2.2026e-02 \\ 
2.0e-02 4.3154e-02 \\ 
4.0e-02 8.2924e-02 \\ 
6.0e-02 1.1969e-01 \\ 
8.0e-02 1.5377e-01 \\ 
1.0e-01 1.8544e-01 \\ 
2.0e-01 3.1521e-01 \\ 
4.0e-01 4.8399e-01 \\ 
6.0e-01 5.8826e-01 \\ 
8.0e-01 6.5865e-01 \\ 
1.0e+00 7.0915e-01 \\ 
			}
            %node [pos=.2,pin={[pin edge={solid,OliveGreen,<-}]90:without capture},inner sep=1pt] {}
            ;
            \addlegendentry{random};

            \addplot [line width = 1.5,color=red]
			table[row sep=crcr]{%
            1.0e-03 1.8418e-03 \\ 
2.5e-03 4.6052e-03 \\ 
5.0e-03 9.1650e-03 \\ 
7.5e-03 1.3625e-02 \\ 
				1.0e-02 1.8410e-02 \\ 
2.0e-02 3.6015e-02 \\ 
4.0e-02 7.0350e-02 \\ 
6.0e-02 1.0248e-01 \\ 
8.0e-02 1.3492e-01 \\ 
1.0e-01 1.6415e-01 \\ 
2.0e-01 2.9677e-01 \\ 
4.0e-01 4.8400e-01 \\ 
6.0e-01 5.8827e-01 \\ 
8.0e-01 6.5866e-01 \\ 
1.0e+00 7.0916e-01 \\ 
			};
            \addlegendentry{hybrid};

            \addplot [mark = x,mark color = red,only marks,mark size = 4]
			table[row sep=crcr]{%
                % 0.01 1.8918e-02 \\
                % 0.04 154.9866 \\
                % 0.1 280.2045 \\
                % 0.4 4.8438e-01 \\
                1 7.0845e-01\\
                .75 6.4190e-01 \\
                .5 5.3972e-01 \\
                .25 3.6618e-01 \\
                .1 1.6360e-01 \\
                .075 1.2526e-01 \\
                .05 8.6097e-02 \\
                .025 4.5714e-02 \\
                .01 1.7447e-02\\
                .0075 1.3254e-02 \\
                .005 8.5065e-03 \\
                .0025 4.5359e-03 \\
			};
            % \addlegendentry{simulation};

%             \addplot [line width = 1.5,dashdotted,color=OliveGreen,forget plot]
% 			table[row sep=crcr]{%
% 				1.0e-02 7.0136e-03 \\ 
% 2.0e-02 2.8808e-02 \\ 
% 4.0e-02 5.4305e-02 \\ 
% 6.0e-02 8.3023e-02 \\ 
% 8.0e-02  1.0850e-01 \\ 
% 1.0e-01 1.4238e-01 \\ 
% 2.0e-01 2.3461e-01 \\ 
% 4.0e-01 3.8825e-01 \\ 
% 6.0e-01 5.0476e-01 \\ 
% 8.0e-01 5.9523e-01 \\ 
% 1.0e+00 6.6495e-01 \\ 
% 			};
%             % \addlegendentry{reactive strategy};
            
%             \addplot [line width = 1.5,color=blue,forget plot]
% 			table[row sep=crcr]{%
% 				1.0e-02 1.8787e-02 \\ 
% 2.0e-02 3.6786e-02 \\ 
% 4.0e-02 7.1161e-02 \\ 
% 6.0e-02 1.0898e-01 \\ 
% 8.0e-02 1.4149e-01 \\ 
% 1.0e-01 1.6036e-01 \\ 
% 2.0e-01 2.7995e-01 \\ 
% 4.0e-01 4.4123e-01 \\ 
% 6.0e-01 5.4659e-01 \\ 
% 8.0e-01 6.2252e-01 \\ 
% 1.0e+00 6.7564e-01 \\ 
% 			}
%             node [pos=.3,pin={[pin edge={solid,OliveGreen,<-}]-90:with capture},inner sep=1pt] {};

%             \addplot [line width = 1.5,dashed,color=red,forget plot]
% 			table[row sep=crcr]{%
% 				1.0e-02 1.8708e-02 \\ 
% 2.0e-02 3.6670e-02 \\ 
% 4.0e-02 7.1143e-02 \\ 
% 6.0e-02 1.0601e-01 \\ 
% 8.0e-02 1.3992e-01 \\ 
% 1.0e-01 1.6093e-01 \\ 
% 2.0e-01 2.8045e-01 \\ 
% 4.0e-01 4.3908e-01 \\ 
% 6.0e-01 5.4465e-01 \\ 
% 8.0e-01 6.1765e-01 \\ 
% 1.0e+00 6.7182e-01 \\ 
% 			};
        \end{axis}
        \end{tikzpicture}}
            \vspace{-.1cm}
            \caption{Average AoII and \gls{MEP} vs. average total number of state changes per slot ($\Ur\bar{q}$) for symmetric processes with energy harvesting rate $\gamma_0 = \gamma_1 = 0.005$. The cross markers represent simulation results.%: i) $\piv = [0~\dots~0 ~1]$, i.e., a device only transmits with full battery, ii) $\piv = \mathbf{1}$, i.e., a device transmits whenever the battery is not depleted, and iii) optimized $\piv$. 
                %Here, $\Ur = 1000$, $\gamma = 0.005$, $\Er = 8$, $\Nr = 100$, $\Rr = 0.8$, and $\sigma^2 = -20$~dB.
                }
            \label{fig:symmetric}
            \vspace{-.55cm}
        \end{figure}

        We also see in Fig.~\ref{fig:symmetric}(a) that the average \gls{AoII} starts decreasing when the processes change state frequently enough. This is because, in this regime, the receiver's estimate is often corrected by a state change (recall the square in Fig.~\ref{fig:AoII_process}) rather than by a successful update. % delivery. %To remove this effect from the observation, we show in Fig.~\ref{fig:avgAoII_ratio} the ratio between the average AoII achieved by the considered strategies and that achieved if no transmission is attempted. We see that the improvement brought by update transmissions becomes smaller as the processes change state more often. 
        % \begin{figure}[t!]
        %     \centering
        %     \input{fig/avgAoII_reduction_vs_transitionRate}
        %        \vspace{-.3cm}
        %     \caption{Ratio between the average AoII achieved by the considered strategies and the average \gls{AoII} achieved if no transmission is attempted vs. the average total number of transitions in a slot ($\Ur\bar{q}$). %: i) $\piv = [0~\dots~0 ~1]$, i.e., a device only transmits with full battery, ii) $\piv = \mathbf{1}$, i.e., a device transmits whenever the battery is not depleted, and iii) optimized $\piv$. 
        %         %Here, $\Ur = 1000$, $\gamma = 0.005$, $\Er = 8$, $\Nr = 100$, $\Rr = 0.8$, and $\sigma^2 = -20$~dB.
        %         }
        %     \label{fig:avgAoII_ratio}
        %     \vspace{-.3cm}
        % \end{figure}

        In Fig.~\ref{fig:symmetric}(b), we plot the \gls{MEP} achieved by the average-\gls{AoII}-optimal strategies examined in Fig.~\ref{fig:symmetric}(a). While the hybrid strategy still slightly outperforms the random strategy, it is noteworthy that the reactive strategy achieves the lowest \gls{MEP}. {This is because when the devices follow the reactive strategy, they do not transmit during the whole period of no state change, and can thus accumulate more energy to report a state change than in the hybrid and random strategies.} %accumulate many energy units for the transmission when there is a state change.} %Therefore, a state change is {reported with higher power than in the hybrid and random strategies.} %successfully even with a single attempt \pp{PP: This is unclear, as the success also depends on the strategies of the other devices and the collisions.}. 
        {Nevertheless, the advantage of the reactive strategy is minor. Fig.~\ref{fig:symmetric} suggests that the hybrid and random strategies achieve a decent performance for both the average \gls{AoII} and the~\gls{MEP}.}
        % \begin{figure}[t!]
        %     \centering
        %     \input{fig/Pmiss_vs_transitionRate}
        %        \vspace{-.3cm}
        %     \caption{The \gls{MEP} achieved by the average-\gls{AoII} minimizing transmission probabilities vs. the average total number of transitions in a slot ($\Ur\bar{q}$). %: i) $\piv = [0~\dots~0 ~1]$, i.e., a device only transmits with full battery, ii) $\piv = \mathbf{1}$, i.e., a device transmits whenever the battery is not depleted, and iii) optimized $\piv$. 
        %         %Here, $\Ur = 1000$, $\gamma = 0.005$, $\Er = 8$, $\Nr = 100$, $\Rr = 0.8$, and $\sigma^2 = -20$~dB.
        %         }
        %     \label{fig:Pmiss}
        %     \vspace{-.3cm}
        % \end{figure}

        \subsection{Asymmetric Processes}
        \begin{figure}[t!]
            \centering
            \subcaptionbox{Average penalty}{
            \begin{tikzpicture}[scale=.9]
		\begin{axis}[%
			width=3.1in,
			height=1in,	
			scale only axis,
			unbounded coords=jump,
			xmin=0.001,
			xmax=1,
            xmode=log,
			xlabel style={font=\color{white!15!black}},
			xlabel={$\Ur \bar{q}$ (transitions/slot)},
			ymode=log,
			ymin=1,
			ymax=2e4,
			yminorticks=true,
			ytick= {1,1e1, 1e2, 1e3, 1e4,1e5},
			ylabel style={yshift=-.1cm,xshift=-.1cm},
			ylabel={average penalty, $\overline{F}$},
			axis background/.style={fill=white},
			title style={font=\bfseries},
			xmajorgrids,
            xminorgrids,
			ymajorgrids,
			yminorgrids,
			legend style={at={(0.17,0)}, legend columns=2,/tikz/every even column/.append style={column sep=.3cm},anchor=south west, legend cell align=left, align=left,draw=none, fill=white, fill opacity=.6,text opacity = 1}
			]

            \addplot [line width = 1.5,dashdotted,color=OliveGreen]
			table[row sep=crcr]{%
            1.0e-03 1.7125e+04 \\ 
2.5e-03 1.0840e+04 \\ 
5.0e-03 8.6749e+03 \\ 
7.5e-03 7.9388e+03 \\ 
2.5e-02 6.5578e+03 \\ 
5.0e-02 3.1020e+03 \\ 
7.5e-02 1.3778e+03 \\ 
1.0e-01 7.7451e+02 \\ 
2.5e-01 1.2345e+02 \\ 
7.5e-01 1.3543e+01 \\ 
1.0e+00 7.5687e+00 \\ 
			};
            \addlegendentry{reactive};
            
			\addplot [line width = 1.5,dashed,color=blue]
			table[row sep=crcr]{%
            1.0e-03 2.5000e+01 \\ 
2.5e-03 5.8992e+01 \\ 
5.0e-03 1.0749e+02 \\ 
7.5e-03 1.4742e+02 \\ 
2.5e-02 2.8574e+02 \\ 
5.0e-02 3.1704e+02 \\ 
7.5e-02 3.0362e+02 \\ 
1.0e-01 2.5156e+02 \\ 
2.5e-01 1.3662e+02 \\ 
7.5e-01 2.7928e+01 \\ 
1.0e+00 2.1325e+01 \\
			}
            %node [pos=.2,pin={[pin edge={solid,black,<-}]90:without capture},inner sep=1pt] {}
            ;
            \addlegendentry{random};
            
            \addplot [line width = 1.5,color=red]
			table[row sep=crcr]{%
            1.0e-03 5.6215e+00 \\ 
2.5e-03 1.3661e+01 \\ 
5.0e-03 2.7095e+01 \\ 
7.5e-03 3.7126e+01 \\ 
2.5e-02 1.1020e+02 \\ 
5.0e-02 1.7107e+02 \\ 
7.5e-02 2.0163e+02 \\ 
1.0e-01 2.1360e+02 \\ 
2.5e-01 1.1822e+02 \\ 
7.5e-01 1.3543e+01 \\ 
1.0e+00 7.5687e+00 \\ 
			};
            \addlegendentry{hybrid};

            \addplot [line width = 1.5,dotted,color=black]
			table[row sep=crcr]{%
            1.0e-03 2.3174e+02 \\ 
2.5e-03 7.5040e+02 \\ 
5.0e-03 1.9554e+03 \\ 
7.5e-03 4.1640e+03 \\ 
1.0e-02 7.1790e+03 \\ 
2.5e-02 1.2416e+04 \\ 
5.0e-02 3.1020e+03 \\ 
7.5e-02 1.3778e+03 \\ 
1.0e-01 7.7451e+02 \\ 
2.5e-01 1.2345e+02 \\ 
5.0e-01 3.0667e+01 \\ 
7.5e-01 1.3543e+01 \\ 
1.0e+00 7.5687e+00 \\ 
			};
            \addlegendentry{average-AoII optimal};

            \addplot [mark = x,mark color = red,only marks,mark size = 4]
			table[row sep=crcr]{%
                % .001 17.4894 \\
                % .0075 148.4259 \\
                .0075 40.2105\\
                .025 111.8602 \\
                .1 216.9484\\
                .25 121.5401 \\
                1 7.6280 \\
			};
            % \addlegendentry{random strategy, simulation};

   %          \addplot [line width = 1.5,color=blue,forget plot]
			% table[row sep=crcr]{%
			% 	0.01 32.8182\\
   %              0.02 61.4184 \\
   %              0.04 110.7188 \\
   %              0.06 165.8437 \\
   %              0.08 202.7469 \\
   %              0.1 212.0990 \\
   %              0.2 292.1408 \\ 
   %              0.4 321.4237 \\
   %              0.6 309.2381 \\
   %              .8 288.6051 \\
   %              1 263.8766 \\
			% }
   %          node [pos=.3,pin={[pin edge={solid,black,<-}]-90:with capture},inner sep=1pt] {};

   %          \addplot [line width = 1.5,dashed,color=red,forget plot]
			% table[row sep=crcr]{%
			% 	0.01 32.2853\\
   %              0.02 61.0747 \\
   %              0.04 110.7188 \\
   %              0.06 156.8622 \\
   %              0.08 201.2042 \\
   %              0.1 211.3085 \\
   %              0.2 289.1327 \\ 
   %              0.4 321.1538 \\
   %              0.6 308.0796 \\
   %              .8 283.8101 \\
   %              1 261.9046 \\
			% };
        \end{axis}
        \end{tikzpicture}}
            \subcaptionbox{\gls{MEP} achieved with the strategies in Fig.~\ref{fig:asymmetric}(a)}{\begin{tikzpicture}[scale=.9]
		\begin{axis}[%
			width=3.1in,
			height=1in,	
			scale only axis,
			unbounded coords=jump,
			xmin=0.001,
			xmax=1,
            xmode=log,
			xlabel style={font=\color{white!15!black}},
			xlabel={$\Ur \bar{q}$ (transitions/slot)},
			ymode=log,
			ymin=1e-3,
			ymax=1,
			yminorticks=true,
			ylabel style={yshift=-.1cm},
			ylabel={\gls{MEP}},
			axis background/.style={fill=white},
			title style={font=\bfseries},
			xmajorgrids,
            xminorgrids,
			ymajorgrids,
			yminorgrids,
			legend style={at={(.99,0.01)}, anchor=south east, legend cell align=left, align=left,draw=none, fill=white, fill opacity=.6,text opacity = 1}
			]

            \addplot [line width = 1.5,dashdotted,color=OliveGreen]
			table[row sep=crcr]{%
            1.0e-03 1.3147e-03 \\ 
2.5e-03 3.2438e-03 \\ 
5.0e-03 6.2846e-03 \\ 
7.5e-03 9.3454e-03 \\ 
2.5e-02 3.0436e-02 \\ 
5.0e-02 1.0000e+00 \\ 
7.5e-02 1.0000e+00 \\ 
1.0e-01 1.0000e+00 \\ 
2.5e-01 1.0000e+00 \\ 
7.5e-01 1.0000e+00 \\ 
1.0e+00 1.0000e+00 \\ 
			};
            \addlegendentry{reactive};
            
			\addplot [line width = 1.5,dashed,color=blue]
			table[row sep=crcr]{%
            1.0e-03 1.4040e-02 \\ 
2.5e-03 3.4383e-02 \\ 
5.0e-03 6.6503e-02 \\ 
7.5e-03 9.6576e-02 \\ 
2.5e-02 2.6320e-01 \\ 
5.0e-02 4.1757e-01 \\ 
7.5e-02 5.1903e-01 \\ 
1.0e-01 5.7009e-01 \\ 
2.5e-01 8.8925e-01 \\ 
7.5e-01 9.9017e-01 \\ 
1.0e+00 9.8442e-01 \\ 
			}
            %node [pos=.2,pin={[pin edge={solid,black,<-}]90:without capture},inner sep=1pt] {}
            ;
            \addlegendentry{random};

            \addplot [line width = 1.5,color=red]
			table[row sep=crcr]{%
            1.0e-03 5.2527e-03 \\ 
2.5e-03 1.3189e-02 \\ 
5.0e-03 2.5777e-02 \\ 
7.5e-03 3.5237e-02 \\ 
2.5e-02 1.1329e-01 \\ 
5.0e-02 2.5081e-01 \\ 
7.5e-02 3.4717e-01 \\ 
1.0e-01 4.7387e-01 \\ 
2.5e-01 9.5770e-01 \\ 
7.5e-01 1.0000e+00 \\ 
1.0e+00 1.0000e+00 \\ 
			};
            \addlegendentry{hybrid};

            \addplot [line width = 1.5,dotted,color=black]
			table[row sep=crcr]{%
            1.0e-03 3.2811e-02 \\ 
2.5e-03 9.1369e-02 \\ 
5.0e-03 1.9416e-01 \\ 
7.5e-03 3.2811e-01 \\ 
1.0e-02 4.7041e-01 \\ 
2.5e-02 1.0000e+00 \\ 
5.0e-02 1.0000e+00 \\ 
7.5e-02 1.0000e+00 \\ 
1.0e-01 1.0000e+00 \\ 
2.5e-01 1.0000e+00 \\ 
5.0e-01 1.0000e+00 \\ 
7.5e-01 1.0000e+00 \\ 
1.0e+00 1.0000e+00 \\ 
			};
            \addlegendentry{average-AoII optimal};

            \addplot [mark = x,mark color = red,only marks,mark size = 4]
			table[row sep=crcr]{%
                % .0075 9.2825e-02 \\
                .0075 3.809e-02 \\
                .025 1.1391e-01 \\
                .1 4.7953e-01 \\
                .25 9.5818e-01 \\
                1 9.9998e-01\\
			};
            % \addlegendentry{simulation};

%             \addplot [line width = 1.5,dashdotted,color=black,forget plot]
% 			table[row sep=crcr]{%
% 				1.0e-02 7.0136e-03 \\ 
% 2.0e-02 2.8808e-02 \\ 
% 4.0e-02 5.4305e-02 \\ 
% 6.0e-02 8.3023e-02 \\ 
% 8.0e-02  1.0850e-01 \\ 
% 1.0e-01 1.4238e-01 \\ 
% 2.0e-01 2.3461e-01 \\ 
% 4.0e-01 3.8825e-01 \\ 
% 6.0e-01 5.0476e-01 \\ 
% 8.0e-01 5.9523e-01 \\ 
% 1.0e+00 6.6495e-01 \\ 
% 			};
%             % \addlegendentry{reactive strategy};
            
%             \addplot [line width = 1.5,color=blue,forget plot]
% 			table[row sep=crcr]{%
% 				1.0e-02 1.8787e-02 \\ 
% 2.0e-02 3.6786e-02 \\ 
% 4.0e-02 7.1161e-02 \\ 
% 6.0e-02 1.0898e-01 \\ 
% 8.0e-02 1.4149e-01 \\ 
% 1.0e-01 1.6036e-01 \\ 
% 2.0e-01 2.7995e-01 \\ 
% 4.0e-01 4.4123e-01 \\ 
% 6.0e-01 5.4659e-01 \\ 
% 8.0e-01 6.2252e-01 \\ 
% 1.0e+00 6.7564e-01 \\ 
% 			}
%             node [pos=.3,pin={[pin edge={solid,black,<-}]-90:with capture},inner sep=1pt] {};

%             \addplot [line width = 1.5,dashed,color=red,forget plot]
% 			table[row sep=crcr]{%
% 				1.0e-02 1.8708e-02 \\ 
% 2.0e-02 3.6670e-02 \\ 
% 4.0e-02 7.1143e-02 \\ 
% 6.0e-02 1.0601e-01 \\ 
% 8.0e-02 1.3992e-01 \\ 
% 1.0e-01 1.6093e-01 \\ 
% 2.0e-01 2.8045e-01 \\ 
% 4.0e-01 4.3908e-01 \\ 
% 6.0e-01 5.4465e-01 \\ 
% 8.0e-01 6.1765e-01 \\ 
% 1.0e+00 6.7182e-01 \\ 
% 			};
        \end{axis}
        \end{tikzpicture}}
            \vspace{-.1cm}
            \caption{Average penalty and \gls{MEP} vs. %the average total number of transitions in a slot (
            $\Ur\bar{q}$ for asymmetric processes with $q_{01}/q_{10} = 0.01$ and energy harvesting rates $\gamma_0 = 0.005$, $\gamma_1 = 0.05$. We consider the power penalty function with $\alpha_0 =  1$ and $\alpha_1 = 2.$ The cross markers represent simulation results. %: i) $\piv = [0~\dots~0 ~1]$, i.e., a device only transmits with full battery, ii) $\piv = \mathbf{1}$, i.e., a device transmits whenever the battery is not depleted, and iii) optimized $\piv$. 
                %Here, $\Ur = 1000$, $\gamma = 0.005$, $\Er = 8$, $\Nr = 100$, $\Rr = 0.8$, and $\sigma^2 = -20$~dB.
                }
            \label{fig:asymmetric}
            \vspace{-.5cm}
        \end{figure}
        
        We now consider asymmetric processes with $q_{01}/q_{10} = 0.01$ and energy harvesting rates $\gamma_0 = 0.005$ and $\gamma_1 = 0.05$, i.e., the critical process state $1$ has a shorter average duration and triggers a higher energy harvesting rate. In this setup, we consider the power penalty function with $\alpha_0 = 1$ and $\alpha_1 = 2$.  Fig.~\ref{fig:asymmetric}(a) depicts the approximate average penalty achieved with the reactive, random, and hybrid strategies with optimized transmission probabilities. We also show simulation results for the hybrid strategy, which closely match the approximation. The performance gain of the hybrid strategy over the random strategy is more pronounced than in the case of symmetric processes. The reactive strategy incurs a high penalty for small transition rates but outperforms the random strategy when $\Ur \bar{q} > 0.2$. Notably, in this regime, the reactive strategy matches the hybrid strategy's performance. This can be explained as follows. When $\bar{q}$ is large, the processes spend relatively short time in state~$1$, and thus pay a smaller penalty than in state~$0$. 
        For example, when $U\bar{q} = 0.25$, if no update is delivered, the processes remain in state $1$ for an average of $79.21$ slots, inducing a penalty of $79.21^2 \approx 6274$ at the end of the period. This penalty is smaller than the penalty of $7921$ incurred at the end of an average state-$0$ period. Therefore, despite the quadratic penalty associated with state $1$, the devices should prioritize reporting transitions to state~$0$. The random strategy fails to make this prioritization. The optimal strategy assigns a high value of $\pi\of{ij}_b$ for $(i,j) = (1,0)$ and a low value for other $(i,j)$, resembling a reactive strategy. 
        For the same reason, transitions to the critical state $1$ are missed with high probability when $\bar{q}$ is large, as seen in Fig.~\ref{fig:asymmetric}(b). One might need to assign a higher penalty power to the critical state. %, we show the \gls{MEP} achieved with the average-penalty-optimal strategies. We observe that when the processes change state frequently enough, all strategies miss the critical state with probability one. This is because the critical state becomes shorter, and thus the quadratic penalty is not effective. 

        Fig.~\ref{fig:asymmetric} also shows the performance of a strategy optimized for the average \gls{AoII}, i.e., $\alpha_0 = \alpha_1 = 1$. Compared to the hybrid approach, this strategy results in a significant increase in both the average penalty and the \gls{MEP} for small process transition rates $\bar{q}$. It also collapses into a reactive strategy at lower $\bar{q}$ values. This highlights that optimizing for a state-independent penalty fails to account for the significance of the critical state estimation error.

        % \hoang{Next steps: 
        % \begin{itemize}
        %     \item Provide results for more general settings: non-binary asymmetric processes, decoding with capture...

        %     \item Explore the tradeoff between average \gls{AoII}, \gls{MEP}, and average \gls{WED}. This can be done by showing the minimum average \gls{AoII} achieved for a given constraint on \gls{MEP} or \gls{WED}.

        %     \item Explore the correlation between the process state transitions and energy harvesting. Find practical scenarios where this is relevant.
        % \end{itemize}

        % The first two points can be done easily. For the last point, we should discuss with Petar.
        % }

				%-----------------------------------------------
				\section{Conclusions} \label{sec:conclusions}
				We studied the state-dependent penalty of outdated status updates received from energy-harvesting devices using the slotted ALOHA protocol. We proposed an efficient and accurate approximation for the average penalty and for the probability of misdetecting a critical period. We conclude that it is important to i) optimally adjust the transmission probabilities to the process state transition and the current battery level to minimize the average penalty, ii) select an appropriate penalty function to capture the varying importance of state estimation errors,  iii) balance the average penalty and the probability of misdetecting a critical period. These insights contribute to the design of effective \gls{IoT} networks under energy and access constraints. {A future direction is to explore feedback from the gateway.}

			\bibliographystyle{IEEEtran}
			\bibliography{IEEEabrv,./biblio}

    % \clearpage
    \appendices

        \section{Process-Battery Profile Evolution of $U-1$ Devices} \label{app:process_battery}
        
        We describe the evolution of the process-battery profile $\Lm$ across slots. Recall that $\ellv' = (\ell'_{(0,0)}, \dots, \ell'_{(0,\Er)}, \ell'_{(1,0)}, \dots, \ell'_{(1,\Er)})$ and $\ellv = (\ell_{(0,0)}, \dots, \ell_{(0,\Er)}, \ell_{(1,0)}, \dots, \ell_{(1,\Er)})$ denote the profiles at the end of two successive slots. Let also $u_{(x',b'),(x,b)}$ be the number of devices whose process-battery state goes from $(x',b')$ to $(x,b)$. % in the transition from $\ellv'$ to $\ellv$. 
    For convenience, we also index $(x,b)$ by its position in %the sequence 
    $\{(0,0), \dots, (0,\Er), (1,0), \dots,(1,\Er)\}$. That is, we also write $\ellv'$ as $(\ell'_1, \dots, \ell'_{2\Er+2})$, $\ellv$ as $(\ell_1, \dots, \ell_{2\Er+2})$, and $u_{(x',b'),(x,b)}$ as $u_{j,k}$ where $j$ and $k$ are the associated indices of $(x',b')$ and $(x,b)$, respectively.
	The transition probabilities of $\Lm$ %between the profiles of the $\Ur-1$ devices 
	are %given by
	\begin{align} \label{eq:trans_profile}
		\P{\ellv' \to \ellv} 
        &= \sum_{\{u_{(x',b'),(x,b)}\}}  \bigg(\prod_{(x',b'),(x,b) \in \{0,1\} \times [0\;:\;\Er]} \notag \\
        &\qquad \qquad \qquad \qquad \P{(x',b') \to (x,b)}^{u_{(x',b'),(x,b)}}\bigg)  \notag\\
		&\qquad \cdot \prod_{j=1}^{2\Er+2} \prod_{k=1}^{2\Er+2} \binom{\ell'_j}{u_{j,1}} \binom{\ell'_j - \sum_{q=1}^{k-1} u_{j,q}}{u_{j,k}}.
	\end{align}
	where the sum is over all values of $\{u_{(x',b'),(x,b)}\}$ such that 
    % \begin{align}
        $u_{(x',b'),(x,b)} \in [0:\min\{\ell'_{(x',b')},\ell_{(x,b)}\}]$, %\\ 
	   $\ell'_{(x',b')} = \textstyle\sum_{(x'',b'')} u_{(x',b'),(x'',b'')}$,  %\\
        and $\ell_{(x,b)} = \textstyle\sum_{(x'',b'')} u_{(x'',b''),(x,b)}$. 
    % \end{align}
	%Furthermore, the steady-state distribution of $\Lm$ is ${\rm Mul}(\Ur-1,2\Er+2,\{\nu_{(x,b)}\}_{x\in\{0,1\}, b\in[0:\Er]})$.

    \section{Transition Probabilities of the Markov Chain $G\of{n}$} 
    \label{app:Gn}

    The transition probabilities is computed using~\eqref{eq:tmp422} %$\P{x, b'\!\to\! b \;|\; \ellv'}$ is a shorthand for $\P{X\of{n} \!=\! x, B\of{n} \!=\! b \;|\; B\of{n-1} \!=\! b', \Lm\of{n-1} \!=\! \ellv'}$, and 
    and a derivation of
	$\P{(x', \hat x', b') \to (x, \hat x, b) \;|\; \ellv'}$ of $G\of{n}$, which we show next. 
	
	\subsubsection{Case 1: $x =x'$, i.e., the process remains in state~$x'$} If $\hat x \ne x' \ne \hat x'$, the device transmits without a successful update or it does not transmit. Therefore,
	\begin{align}
		&\P{(x', \hat x', b') \to (x', \hat x, b) \;|\; \ellv'} \notag \\
		&= q_{x'x'}\big[ \pi_{b'}\of{x'x'} (1-\omega_{b',\ellv'})\phi\sups{trans.}_{b} +  (1-\pi_{b'}\of{x'x'})\phi\sups{no~trans.}_{b'\to b}\big].
	\end{align}
	If $\hat x = x' = \hat x'$, the receiver's estimate remains correct regardless of the transmission of the device, and thus
		\begin{multline}
		\P{(x', x', b') \to (x', x', b) \;|\; \ellv'} \\= q_{x'x'}\big[ \pi_{b'}\of{x'x'} \phi\sups{trans.}_{b} +  (1-\pi_{b'}\of{x'x'})\phi\sups{no~trans.}_{b'\to b}\big].
	\end{multline} 
	If $\hat x = x' \ne \hat x'$, the receiver successfully delivers an update to the gateway, and thus
	\begin{align}
		\P{(x', \hat x', b') \to (x', x', b) \;|\; \ellv'} &= q_{x'x'} \pi_{b'}\of{x'x'} \omega_{b',\ellv'}\phi\sups{trans.}_{b}, \notag  \\ &\qquad \forall \hat x' \ne x'.
	\end{align}
	Finally, the event $\hat x \ne x' = \hat x'$ cannot occur, i.e., 
	% \begin{equation}
		$\P{(x', x', b') \to (x', \hat x, b) \;|\; \ellv'} = 0$ for $\hat x \ne x'$.
	% \end{equation}
    Therefore, we obtain~\eqref{eq:PtransG_x=x'}.

	\subsubsection{Case 2: $x \ne x'$, i.e., the process changes state} Using similar arguments of as the previous case for different relations of $\hat x$, $x'$, and $\hat x'$, we obtain~\eqref{eq:PtransG_x_ne_x'}.

    %---------------------
     \section{Proof of Theorem~\ref{th:avg_penalty}} \label{proof:avg_penalty}

            %For a penalty function $p$, 
        We compute the average penalty~\eqref{eq:avg_penalty} as 
		%An \gls{AoII} metric of the form $F(\Lambda) =   $ is derived as
		\begin{align}
			\overline{F} %&= \lim\limits_{k \to \infty} \frac{1}{k} \sum_{n=1}^k f_{X\of{n}}(\Lambda(n)) \\ 
            &= \lim_{m \to \infty} \frac{\sum_{i=1}^m \sum_{j = n_{i}}^{n_{i+1}-1} f_{X\of{j}}(\Lambda(j))}{\sum_{i=1}^m (W_i + Y_i)} \label{eq:tmp766} \\
			&= \lim_{m \to \infty} \frac{\frac{1}{m}\sum_{i=1}^m %(Y\of{i} p_{\tilde{X}\of{i}}(0) + 
            \sum_{j=1}^{W_i} f_{{X}\of{j}}(j)}{\frac{1}{m}\sum_{i=1}^m (W_i + Y_i)}.  \label{eq:tmp767}
        \end{align}
        In~\eqref{eq:tmp766}, we break the time horizon into $m$ durations, each containing a \gls{WED} followed by a \gls{CED}, and denote the first time slot of the $i$th duration by $n_i$. Equation~\eqref{eq:tmp767} follows because the penalty is positive only within a wrong-estimate period. 
        We further expand the the numerator in~\eqref{eq:tmp767} as
        \begin{align}
			%&\sum_{y=1}^\infty %\frac{|\{i\in [m] \colon Y_i = y\}|}{m} y f(0) \notag \\
			%&\qquad \quad  + 
            &\frac{1}{m}\sum_{w=1}^\infty \sum_{x=0}^1 |\{i\in [m] \colon W\of{i} = w, \tilde{X}\of{i} = x\}| \sum_{j=1}^{w} f_{x}(j) \notag \\
            &= \sum_{x=0}^1 \frac{|\{i\in [m] \colon \tilde{X}\of{i} = x\}|}{m}  \notag \\
            &\quad \cdot \sum_{w=1}^\infty  \frac{|\{i\in [m] \colon W\of{i} = w, \tilde{X}\of{i} = x\}|}{|\{i\in [m] \colon \tilde{X}\of{i} = x\}|}  \sum_{j=1}^{w} f_{x}(j) % \\
			% &= \frac{\E{Y} f(0) + \E{\sum_{j=1}^{W} f(j)}}{\E{W} + \E{Y}}. \label{eq:tmp637}
		\end{align}
        where we recall that $\tilde{X}\of{i}$ denotes the process state during the $i$th wrong-estimate period. As $m \to\infty$, $\frac{|\{i\in [m] \colon \tilde{X}\of{i} = x\}|}{m}$ converges to $\P{\tilde{X} = x}$, and $\frac{|\{i\in [m] \colon W\of{i} = w, \tilde{X}\of{i} = x\}|}{|\{i\in [m] \colon \tilde{X}\of{i} = x\}|}$ to $\P{W_x = w}$. Furthermore, the denominator in~\eqref{eq:tmp767} converges to $\E{W} + \E{Y}$. Therefore, we obtain $\overline{F} = \frac{\Gamma}{\E{W} + \E{Y}}$.
        
		% For the average \gls{AoII}, $f_x(j) = j$, and we have that $\sum_{j=1}^{W_x} f_x(j) = \frac{1}{2}(W_x + W_x^2)$ for both $x = 0$ and $x = 1$. From this, we obtain~\eqref{eq:avgAoII_WY}. 

        %For the linear penalty function $f_x(j) = \alpha_x j$, we have that $\sum_{j=1}^{W_x} f_x(j) = \frac{\alpha_x}{2}(W_x + W_x^2)$. From this, we obtain~\eqref{eq:pen_linear_WY}. The average \gls{AoII} follows by taking $\alpha_0 = \alpha_1 = 1$.
        
        For the power penalty function $f_x(j) = j^{\alpha_x}$, if $\alpha_x$ is a nonnegative integer, we obtain using Faulhaber's formula~\cite{knuth1993johann} 
        that 
        $\sum_{j=1}^{W_x} f_x(j) = \frac{1}{\alpha_x + 1} \sum_{k=0}^{\alpha_x} \binom{\alpha_x + 1}{k} \Br_{k} W_x^{\alpha_x-k + 1}$,
        which leads to~\eqref{eq:pen_power_WY}. The average \gls{AoII} follows by taking $\alpha_0 = \alpha_1 = 1$.
        
        %For $f_x(j) = e^{\alpha_x j} - 1$, we obtain after some simple manipulations that $\sum_{j=1}^{W_x} f_x(j) = \frac{e^{\alpha_x}}{e^{\alpha_x} - 1} (e^{\alpha_xW_x} - 1) - W_x$, which leads to~\eqref{eq:pen_exp_WY}.
        %For the \gls{AoII} violation probability, $f(j) = \ind{j>\theta}$, and we have that $\sum_{j=1}^{W} f(j) = (W-\theta)^+$. Indeed, within an error duration of length $W$, the AoI exceeds $\theta$ in the last $(W-\theta)^+$ slots. Substituting this into~\eqref{eq:tmp637}, we obtain~\eqref{eq:AoIIVP_YW}.

    %-------------------------
    \section{Approximate \gls{CED} Distribution}
    \label{app:approx_CED}
    %\subsection{Approximate \gls{CED} Distribution}\label{sec:approxAoI_Y}

            					\begin{figure}[t!]
			\centering
						\vspace{-.2cm}
			\scalebox{.64}{\begin{tikzpicture}
					\tikzset{node distance=3cm, % Minimum distance between two nodes. Change if necessary.
						every state/.style={ % Sets the properties for each state
							semithick,
							fill=gray!10},
						initial text={},     % No label on start arrow
						double distance=4pt, % Adjust appearance of accept states
						every edge/.style={  % Sets the properties for each transition
							draw,
							->,>=stealth',     % Makes edges directed with bold arrowheads
							auto,
							semithick}}
					\node[state] (b0) {$1,1,0$};
					\node[state, right of= b0] (b1) {$1,1,1$};
					\node[right of= b1] (mid) {$\dots$};
					\node[state, right of= mid,align=center] (b3) {$1,1$, \\ $\Er\!-\!1$};
					\node[state, right of= b3] (b4) {$1,1,\Er$};
					
					\node[state, above=1.5cm of mid] (A) {$\Ar\sub{WE}$};
					
					\draw (b0) edge[loop above] node[above,midway] {} (b0);
					\draw (b1) edge[loop above] node[above,midway] {} (b1);
					\draw (b3) edge[loop above] node[above,midway] {} (b3);
					\draw (b4) edge[loop above] node[above,midway] {} (b4);
					
					\draw (b0) edge[bend left=20] node[above,midway] {} (b1);
					\draw (b1) edge[bend left=20] node[above,midway] {} (mid);
					\draw (mid) edge[bend left=20] node[above,midway] {} (b3);
					\draw (b3) edge[bend left=20] node[above,pos=.5] {} (b4);
					
					\draw (b1) edge[bend left=18] node[above=-.07cm,midway] {} (b0);
					\draw (mid) edge[bend left=21] node[above,midway] {} (b0);
					\draw (b3) edge[bend left=24] node[above=-.07cm,pos=.54,rotate=-2] {} (b0);
					\draw (b4) edge[bend left=27] node[above=-.05cm,pos=.5] {} (b0);
					
					%				\draw (b1) edge[bend left=15] node[above=-.07cm,midway] {$\rho_1$} (b1);
					\draw (mid) edge[bend left=18] node[above,midway] {} (b1);
					\draw (b3) edge[bend left=21] node[above=-.07cm,pos=.5] {} (b1);
					\draw (b4) edge[bend left=24] node[above=-.07cm,pos=.5] {} (b1);
					
					\draw[color=red] (b0) edge[bend left=12] node[above=-.07cm,midway] {} (A);
					\draw[color=red] (b1) edge[bend left=10] node[above=-.07cm,midway] {} (A);
					\draw[color=red] (mid) edge node[above,midway] {} (A);
					\draw[color=red] (b3) edge[bend right=10] node[above=-.07cm,pos=.54,rotate=-2] {} (A);
					\draw[color=red] (b4) edge[bend right=12] node[above=-.05cm,pos=.5] {} (A);

                    %%%%%%%%%%%%%%%
					\node[state, above=3.1cm of b0] (b10) {$0,0,0$};
					\node[state, right of= b10] (b11) {$0,0,1$};
					\node[right of= b11] (mid1) {$\dots$};
					\node[state, right of= mid1,align=center] (b13) {$0,0$, \\ $\Er\!-\!1$};
					\node[state, right of= b13] (b14) {$0,0,\Er$};
					
					% \node[state, below of= b0] (A) {$\Ar_{\setminus(x,\hat x)}$};

					\draw (b10) edge[loop below] node[above,midway] {} (b10);
					\draw (b11) edge[loop below] node[above,midway] {} (b11);
					\draw (b13) edge[loop below] node[above,midway] {} (b13);
					\draw (b14) edge[loop below] node[above,midway] {} (b14);
					
					\draw (b10) edge[bend right=20] node[above,midway] {} (b11);
					\draw (b11) edge[bend right=20] node[above,midway] {} (mid1);
					\draw (mid1) edge[bend right=20] node[above,midway] {} (b13);
					\draw (b13) edge[bend right=20] node[above,pos=.5] {} (b14);
					
					\draw (b11) edge[bend right=18] node[above=-.07cm,midway] {} (b10);
					\draw (mid1) edge[bend right=21] node[above,midway] {} (b10);
					\draw (b13) edge[bend right=24] node[above=-.07cm,pos=.54,rotate=-2] {} (b10);
					\draw (b14) edge[bend right=27] node[above=-.05cm,pos=.5] {} (b10);
					
					%				\draw (b1) edge[bend left=15] node[above=-.07cm,midway] {$\rho_1$} (b1);
					\draw (mid1) edge[bend right=18] node[above,midway] {} (b11);
					\draw (b13) edge[bend right=21] node[above=-.07cm,pos=.5] {} (b11);
					\draw (b14) edge[bend right=24] node[above=-.07cm,pos=.5] {} (b11);

                    \draw[color=red] (b10) edge[bend right=12] node[above=-.07cm,midway] {} (A);
					\draw[color=red] (b11) edge[bend right=10] node[above=-.07cm,midway] {} (A);
					\draw[color=red] (mid1) edge node[above,midway] {} (A);
					\draw[color=red] (b13) edge[bend left=10] node[above=-.07cm,pos=.54,rotate=-2] {} (A);
					\draw[color=red] (b14) edge[bend left=12] node[above=-.05cm,pos=.5] {} (A);

                    %%%%%%%%%%%%%%%%
                    \draw[gray] (b11) edge[in=65,out=-130] (b0);
                    \draw[gray] (b11) edge[bend right=20] (b1);
                    \draw[gray] (mid1) edge[bend right=10] (b0);
                    \draw[gray] (mid1) edge[in=63,out=-130] (b1);
                    \draw[gray] (b13) edge[bend right=10] (b0);
                    \draw[gray] (b13) edge[bend left=18] (b1);
                    \draw[gray] (b14) edge[in=30,out=-115] (b0);
                    \draw[gray] (b14) edge[bend left=10] (b1);

                    \draw[gray] (b1) edge[in=-65,out=130] (b10);
                    \draw[gray] (b1) edge[bend right=20] (b11);
                    \draw[gray] (mid) edge[bend left=10] (b10);
                    \draw[gray] (mid) edge[in=-63,out=130] (b11);
                    \draw[gray] (b3) edge[bend left=10] (b10);
                    \draw[gray] (b3) edge[bend right=18] (b11);
                    \draw[gray] (b4) edge[in=-30,out=115] (b10);
                    \draw[gray] (b4) edge[bend right=10] (b11);
                    
				\end{tikzpicture}
			}

            \vspace{-.2cm}
			\caption{Markov chain $M\sub{CE}$ describing the slot-wise evolution of a device within a correct-estimate period, i.e., when $(X,\hat X) \in \{(0,0),(1,1)\}$. The absorbing state $\Ar\sub{WE}$ %indicates the first slot where the receiver's estimate becomes wrong. It 
            represents all states of the chain $(X\of{n},\hat X\of{n}, B\of{n})$ where $X\of{n} \ne \hat X\of{n}$.}
			\label{fig:markov_CED}
						\vspace{-.4cm}
		\end{figure}

            Similar to the derivation of the WED distribution, we modify the Markov chain $(X\of{n}, X\of{n}, B\of{n})$ to obtain another chain that describes the slot-wise evolution of a device within a correct-estimate period. Specifically, we combine all states $(x,\hat x,b)$ with $x \ne \hat x$ and $b \in [0:\Er]$ into a single state $\Ar\sub{WE}$ that represents all slots with a wrong estimate. % an \gls{AoI} refresh. In other words, we redirect all transitions that lead to an \gls{AoI} refresh into a new state $\refresh''$.    
		%Furthermore, we split the state $b \in [\Er]$ in $M_1$ into two states $\Sc_b$ and $\fail,0$, specifying whether the latest update of the device has or has not reached the gateway, respectively, before the current slot. 
		%Note that if the device is in state $\Sc_i$, $i\in [\Er]$, a retransmission of the latest update cannot trigger an \gls{AoI} refresh. 
		We call the resulting terminating Markov chain $M\sub{CE}$ and depict it in Fig.~\ref{fig:markov_CED}. 
		We denote by $\Ts\sub{CE}$ the matrix containing the probabilities of transitions between the transient states and $\av\sub{CE}$ the vector containing the probabilities of transitions from the transient states to the absorbing state, obtained using~\eqref{eq:ptrans_simp_chain_x_=_x'} and~\eqref{eq:ptrans_simp_chain_x_ne_x'}. %We obtain $\Ts\sub{WE}$ and $\av\sub{WE}$ using~\eqref{eq:ptrans_simp_chain_x_=_x'} and~\eqref{eq:ptrans_simp_chain_x_ne_x'}. 
        Following similar steps as in the previous subsection, we obtain the \gls{PMF} and mean of $Y$ as
			\begin{align} 
                \P{Y = y} &= \tauv\sub{CE}^\T \Ts\sub{CE}^{y-1}\av\sub{CE}], \quad y = 1,2,\dots \label{eq:approx_dist_Y} \\
			         \E{Y} &=  \tauv\sub{CE}^\T [(\Is-\Ts\sub{CE})^{-1} \mathbf{1}], \label{eq:approx_EY}
			\end{align}
        where $\tauv\sub{CE} = (\tau_{(0,0,0)}, \dots, \tau_{(0,0,\Er)},\tau_{(1,1,0)}, \dots, \tau_{(1,1,\Er)})$ with 
        \begin{equation}
            \tau_{(x,\hat x,b)} = \frac{\P{\to (x,\hat x,b)}}{\sum_{(x',\hat x') \in \{(0,0),(1,1)\}, b' \in [0:\Er]} \P{\to (x',\hat x',b')}}
        \end{equation}
        for $x = \hat x$.
			
			\section{Approximate \gls{MEP}} \label{app:approx_MEP}
			
						\begin{figure}[t!]
				\centering
				%			\vspace{-.5cm}
				\scalebox{.64}{\begin{tikzpicture}
						\tikzset{node distance=3cm, % Minimum distance between two nodes. Change if necessary.
							every state/.style={ % Sets the properties for each state
								semithick,
								fill=gray!10},
							initial text={},     % No label on start arrow
							double distance=4pt, % Adjust appearance of accept states
							every edge/.style={  % Sets the properties for each transition
								draw,
								->,>=stealth',     % Makes edges directed with bold arrowheads
								auto,
								semithick}}
						\node[state] (b0) {$1,0,0$};
						\node[state, right of= b0] (b1) {$1,0,1$};
						\node[right of= b1] (mid) {$\dots$};
						\node[state, right of= mid,align=center] (b3) {$1,0$, \\ $\Er\!-\!1$};
						\node[state, right of= b3] (b4) {$1,0,\Er$};
						
						\node[state, above = 1.2cm of b0] (n00) {$0,0$};
						\node[state, above = 1.2cm of b4] (n11) {$1,1$};
						
						\draw (b0) edge[loop above] node[above,midway] {} (b0);
						\draw (b1) edge[loop above] node[above,midway] {} (b1);
						\draw (b3) edge[loop above] node[above,midway] {} (b3);
						\draw (b4) edge[loop above] node[above,midway] {} (b4);
						
						\draw (b0) edge[bend left=20] node[above,midway] {} (b1);
						\draw (b1) edge[bend left=20] node[above,midway] {} (mid);
						\draw (mid) edge[bend left=20] node[above,midway] {} (b3);
						\draw (b3) edge[bend left=20] node[above,pos=.5] {} (b4);
						
						\draw (b1) edge[bend left=18] node[above=-.07cm,midway] {} (b0);
						\draw (mid) edge[bend left=21] node[above,midway] {} (b0);
						\draw (b3) edge[bend left=24] node[above=-.07cm,pos=.54,rotate=-2] {} (b0);
						\draw (b4) edge[bend left=27] node[above=-.05cm,pos=.5] {} (b0);
						
						%				\draw (b1) edge[bend left=15] node[above=-.07cm,midway] {$\rho_1$} (b1);
						\draw (mid) edge[bend left=18] node[above,midway] {} (b1);
						\draw (b3) edge[bend left=21] node[above=-.07cm,pos=.5] {} (b1);
						\draw (b4) edge[bend left=24] node[above=-.07cm,pos=.5] {} (b1);
						
						\draw[color=red] (b0) edge[bend right=30] node[above=-.07cm,midway] {} (n00);
						\draw[color=red] (b1) edge[bend right=14] node[above=-.07cm,midway] {} (n00);
						\draw[color=red] (mid) edge[bend right=16] node[above,midway] {} (n00);
						\draw[color=red] (b3) edge[bend right=18] node[above=-.07cm,pos=.54,rotate=-2] {} (n00);
						\draw[color=red] (b4) edge[bend right=20] node[above=-.05cm,pos=.5] {} (n00);
						
						\draw[color=blue] (b4) edge[bend left=30] node[above=-.07cm,midway] {} (n11);
						\draw[color=blue] (b3) edge[bend left=14] node[above=-.07cm,midway] {} (n11);
						\draw[color=blue] (mid) edge[bend left=16] node[above,midway] {} (n11);
						\draw[color=blue] (b1) edge[bend left=18] node[above=-.07cm,pos=.54,rotate=-2] {} (n11);
						\draw[color=blue] (b0) edge[bend left=20] node[above=-.05cm,pos=.5] {} (n11);
					\end{tikzpicture}
				}

                \vspace{-.2cm}
				\caption{Markov chain $M\sub{ME}$ describing i) the slot-wise battery level evolution of a device when the process state and its estimate are $(X,\hat X) = (1,0)$, and ii) the transition to any other state of $(X,\hat X)$. %The states $(0,0)$ and $(1,1)$ represent all states of the chain $(X\of{n},\hat X\of{n},B\of{n})$ where $(X,\hat X)$ equals $(0,0)$ and $(1,1)$, respectively.
                }
				\label{fig:markov_miss}
							\vspace{-.4cm}
			\end{figure}
   
            %Using an abrobing Markov chain analysis, 
            We now derive the \gls{MEP} $\Pmiss$ under Simplification~\ref{simplification}. A transition to the critical state $1$ is missed by the receiver if: i) the transition of the process to state $1$ {in a slot} is not notified to the receiver {in the same slot}, and ii) no update is delivered for the whole period over which the process remains in state~$1$. We denote by $\Punnoticed{b}$ the probability that event (i) occurs and the device ends up having battery level $b$, and by $\Poblivious{b}$ the probability that event (ii) occurs given that event (i) has occurred and the device has battery level $b$ after that. We next compute these probabilities.

            First, $\Punnoticed{b}$ is the probability that, given that the process moves from state~$0$ to state~$1$ and the device's battery level moves from some value $b'$ to $b$, the receiver's estimate remains at $0$. We can compute it as
			\begin{align}
				\!\!\!\Punnoticed{b} &=  \bigg(\sum_{b' \in [0:\Er]} p_{(0,0,b')}\bigg)^{\!\!-1} %\notag \\&\quad \cdot
                \!\!\!\sum_{b' \in [0:\Er]} \!p_{(0,0,b')}\big[\pi_{b'}\of{01} (1-\bar{\omega}_{b'}) \phi\sups{trans.}_b \notag \\
                &\qquad\qquad \qquad \qquad \qquad \quad  + (1-\pi\of{01}_{b'}) \phi\sups{no~trans.}_{b' \to b} \big].
			\end{align}
            We let $\rhov = (\Punnoticed{0}, \dots, \Punnoticed{\Er}).$
            
            To compute $\Poblivious{b}$, note that during a period where the receiver is not notified of the critical event, we have that $(X,\hat X) = (1,0)$. The battery evolution of the device within this period is described by the Markov chain $M\sub{ME}$ depicted in Fig.~\ref{fig:markov_miss}. To obtain this chain, we combine all states of the chain $(X\of{n},\hat X\of{n},B\of{n})$ where $(X,\hat X)$ is given by $(0,0)$ and $(1,1)$, and we ignore the states where $(X,\hat X) = (0,1)$. The chain $M\sub{ME}$ is an absorbing Markov chain with two absorbing states, namely, $(0,0)$ and $(1,1)$. Note that $\Poblivious{b}$ is the probability that the chain is absorbed into the state $(0,0)$ (rather than $(1,1)$) given that it starts in the transient state $(1,0,b)$. According to %~\cite[Th.~3.3.7]{Kemeny1976_Markov},
            the property of absorbing Markov chains,
            the vector $\kappav = (\Poblivious{0}, \dots,\Poblivious{
            \Er})$ is given by
			\begin{align}
				\kappav = (\Is - \Ts\sub{miss})^{-1} {\av}\sub{miss},
			\end{align}
            where $\Ts\sub{miss}$ contains the probabilities of transitions between the transient states and ${\av}\sub{miss}$ contains the probabilities of transitions from the transient states to the absorbing state $(0,0)$ of the chain $M\sub{ME}$, obtained using~\eqref{eq:ptrans_simp_chain_x_=_x'} and~\eqref{eq:ptrans_simp_chain_x_ne_x'}.
            
			Finally, using the law of total probability, we compute the \gls{MEP} as
            % \begin{align}
                $\Pmiss = \rhov^\T \kappav$.
            % \end{align}
		\end{document}